\documentclass{jaa}
\usepackage{natbib}
\usepackage{graphicx}
\usepackage{float}
\usepackage{color}

\begin{document}\sloppy

\title{Atmospheric parameters and chemical abundances of \\
the A-type eclipsing binary system RR Lyncis A and B}


\author{Yoichi Takeda\textsuperscript{1}}
\affilOne{\textsuperscript{1}11-2 Enomachi, Naka-ku, Hiroshima-shi, 730-0851, Japan}


\twocolumn[{

\maketitle

\corres{ytakeda@js2.so-net.ne.jp}

\msinfo{}{}

\begin{abstract}
A spectroscopic study was carried out for the double-line A-type eclipsing
binary system RR~Lyn A+B based on the disentangled spectra, with an aim of 
clarifying the differences in photospheric chemical compositions between 
the components, where $T_{\rm eff}$ (effective temperature) and $v_{\rm t}$ 
(microturbulence) were determined from Fe lines. 
The resulting abundances of 30 elements revealed the following characteristics. 
(1) The brighter/hotter A shows metal-rich trends of classical Am stars; i.e., 
heavier elements generally show overabundances tending to increase towards higher 
$Z$ (atomic number) with exceptionally large deficit of Sc, while light elements 
such as CNO show underabundances. 
(2) Meanwhile, the abundances of fainter/cooler B are closer to the solar 
composition ([X/H]~$\sim 0$ for intermediate $Z$ elements such as Fe group)
though [X/H] does exhibit a slightly increasing tendency with $Z$, which 
suggests that B is a kind of marginal Am star with almost normal metallicity.   
This consequence is in contrast to the results of previous studies, 
which reported B to be of metal-deficient nature similar to $\lambda$ Boo stars.  
Such distinctions of chemical abundances between A and B may serve as a key 
to understanding the condition for the emergence of Am phenomenon.
\end{abstract} 


}]


\doinum{}
\artcitid{\#\#\#\#}
\volnum{000}
\year{0000}
\pgrange{1--}
\setcounter{page}{1}
\lp{1}

\section{Introduction} 

It is a long-standing problem which has been attracted stellar spectroscopists 
that a fraction of A- and late B-type stars on the upper main sequence show 
various kinds of compositional anomalies. Although the basic classification scheme 
of these chemically peculiar stars of diversified characteristics (e.g., Si stars, 
SrCrEu stars, Am stars, HgMn stars, $\lambda$~Boo stars, etc.) was almost established 
about a half century ago (e.g., Preston 1974), our understanding on the nature and 
origin of their chemical anomalies is still far from satisfactory.

One of the key approaches to clarify the condition for the emergence of chemical 
peculiarities is to study and compare the abundances of the components of
double-line spectroscopic binaries, because both should have had the same compositions
when they were born. In particular, eclipsing binaries would be preferable, 
since their stellar parameters are more precisely determinable.
Yet, chemical abundance determinations of double-line binaries generally 
suffer considerable difficulties, because  spectral lines are so intricately
mixed that reliable measurements of line strengths for both components on the
composite spectra are not easy to practice.

Fortunately, the advent of the efficient spectrum-disentangling technique, which can 
numerically reproduce the spectra of both components as resolved (thanks to the improved 
computational power and availability of high-S/N and high-resolution spectra nowadays), 
changed this situation dramatically.

Takeda et al. (2019; hereinafter referred to as Paper~I) applied this method
to the spectra of five eclipsing binaries (AR~Aur, $\beta$~Aur, WW~Aur, YZ~Cas,
and RR~Lyn) covering various phases to obtain the disentangled spectra in five 
selected regions. Then, the chemical abundances of several elements (with the 
main attention being paid to light elements such as CNO) were determined  
for both components in order to discuss the compositional differences between them.

Especially important is a system of two stars with rather similar parameters,
where one shows a distinct type of chemical peculiarity while the other does not.
AR~Aur corresponds to such an interesting case; i.e., the primary (A) is a typical 
HgMn star but such an anomaly is not seen in the secondary (B), despite that both 
are similar late B-type stars. Accordingly, the author recently carried out 
an extensive reanalysis of this binary system by fully exploiting the disentangled 
spectra of wide wavelength ranges (Takeda 2025; hereinafter Paper~II), where the 
stellar parameters [$T_{\rm eff}$ (effective temperature) and $v_{\rm t}$ 
(microturbulence)] were spectroscopically determined from many Fe lines and the 
abundances of 28 elements were derived (the non-LTE effect was taken into account 
for 10 elements). The typical characteristics of HgMn peculiarities with large 
dispersion (very deficient N, Al, Sc, Ni or very overabundant P, Mn) were then 
confirmed in A, while B was found to show a comparatively weak and rather organized 
peculiarity (almost linearly increasing with the atomic number $Z$).  
It is interesting that such remarkably dissimilar types of chemical peculiarities 
are observed in A and B (with only a small $T_{\rm eff}$ difference of $\sim$~500~K). 

RR~Lyn is also counted as such an interesting eclipsing binary system (orbital period 
is $P = 9.95$~d) consisting of two late~A-type stars ($T_{\rm eff}$ differs 
by at least several hundred K), where the brighter A is a Am star of
metal-rich character while the chemical composition of the fainter B is distinctly 
different (reported to be even metal-poor). Therefore, it is worthwhile to 
investigate the differences of physical properties (in terms of stellar parameters
and chemical abundances) between RR~Lyn A and B as precisely as possible, which may 
provide us with a key to understanding the condition for the advent of Am phenomenon.    

Unfortunately, despite that not a few investigators carried out spectroscopic studies
on this binary system as briefly summarized in Table~1, the results reported in these 
previous publications are diversified and unconformable, from which the problematic 
points may be identified as follows.  
\begin{itemize}
\item
Since the luminosity of B is appreciably lower ($\sim~$25--30\%) than that of A,
line strengths of B appear much weaker than those of A due to the dilution effect, 
which means that measuring B's spectral lines in the composite spectra is a difficult
task. Actually, studies of RR~Lyn in the old days based on data of low-quality 
photographic spectrograms had to regard RR~Lyn A+B as if a single star ($\sim$~A), 
disregarding the lines of B. Most of the other investigations laboriously measured the 
lines of A and B separately in the A+B mixed spectra, though the number of available lines
was comparatively limited in this approach. It is only the analysis of Paper~I that 
employed disentangled spectra for this star. 
\item
Thanks to the merit of eclipsing binary, $\log g$ (surface gravity) values for A and B 
are precisely determined in this system. However, $T_{\rm eff}$ and $v_{\rm t}$ values
adopted in previous studies, which play significant roles in abundance determinations, 
are considerably diversified ($T_{\rm eff}$ differences amounting to $\lesssim$~600--700~K
and $v_{\rm t}$ being discrepant by $\lesssim$~50--100\%) as can be recognized in Table~1. 
These parameters have to be more reliably established by all means; most preferably 
by the spectroscopic method using many lines. 
\item
Almost all previous investigations concluded that the brighter A is metal-rich (though 
its extent appreciably differs from each other), which is reasonably understandable 
because A is believed to be a typical Am star. Meanwhile, it is rather puzzling that 
the fainter B was reported to be mildly metal-poor ([Fe/H] is $\sim -0.2$ to $\sim -0.4$) 
in most of the past studies.  Such a subsolar metallicity is hard to understand for
a comparatively young population~I star. Admittedly, there is a group of chemically
peculiar stars (so-called $\lambda$~Boo stars) which show deficiencies in refractory 
Fe group elements (presumably caused by dust--gas separation mechanism) while volatile 
elements (such as CNO) remain almost normal (see, e.g., Venn \& Lambert 1990). 
Actually, Jeong et al. (2017) affirmed that RR~Lyn~B is a $\lambda$~Boo star.
It should be noted, however, that $\lambda$~Boo stars are generally known to be rapid 
rotators. Since both RR~Lyn A and B are evidently slow rotators, this argument needs 
to be viewed with caution. It is thus of primary importance to check (by precise 
determinations of [Fe/H] based on many lines) whether or not B is really underabundant 
in metals.   
\end{itemize}

These problems motivated the author to carry out a new spectroscopic study on 
the RR~Lyn A+B system in a similar manner to that adopted for AR~Aur (Paper~II),
which is a refined and extended analysis (compared to that of Paper~I)  
where the main attention is paid to the following points.   
\begin{itemize}
\item[--]
Disentangled spectra are obtained for as wide wavelength regions as possible
(from violet to near infrared), which makes measurements of equivalent widths 
for a large number of spectral lines ($\sim 1000$) feasible.
\item[--]
In contrast to Paper~I, where catalogue values were assigned to $T_{\rm eff}$   
(which are originally the photometric determinations of Khaliullin e al. 2001) 
and Takeda et al.'s (2008) empirical $T_{\rm eff}$-dependent formula was
employed for $v_{\rm t}$, ($T_{\rm eff}$, $v_{\rm t}$) are spectroscopically
established from a number of Fe lines by requiring the consistency of
abundances (as done in Paper~II).   
\item[--]
Based on such determined atmospheric parameters, chemical abundances of many 
elements are derived (mainly from the measured equivalent width or 
additionally by using the spectrum-fitting technique), where the non-LTE
is taken into account wherever possible (as in Paper~II). 
\end{itemize}
  
\setcounter{table}{0}
\begin{table*}[h]
\begin{minipage}{160mm}
\begin{center}
\caption{Stellar parameters of RR~Lyn A and B adopted in previous investigations.}
\small
\begin{tabular}{c c@{ }c@{ }c@{ }c c@{ }c@{ }c@{ }c}
\hline\hline
Authors & $T_{\rm eff,A}$ & $\log g_{\rm A}$ & $v_{\rm t,A}$ & [Fe/H]$_{\rm A}$ & 
$T_{\rm eff,B}$ & $\log g_{\rm B}$ & $v_{\rm t,B}$ & [Fe/H]$_{\rm B}$ \\
   &   (K)  &  (dex) & (km~s$^{-1}$) & (dex) &   (K)  &  (dex) & (km~s$^{-1}$) & (dex) \\
\hline
Smith (1971)                         &  (8125) &       &   (7.0) & (6.74)$^{*}$ &       &       &       &          \\
Rachkovskaya (1974)                  &  (7750)$^{\#}$ &       &   (4.7) &           &       &       &       &          \\
Kondo (1976)                         &  8100   & 3.91  &   7.0   &  7.32$^{*}$  &  6900 & 4.07  &  3.0  & 6.54$^{*}$ \\
Burkhart \& Coupry (1991)            &  7960   &       &         &   +0.35   &  7500 &       &       & $-0.3$ \\
Lyubimkov \& Rachkovskaya (1995b)    &  (7850) &(3.95) &   (5.8) &   (+0.11) &       &       &       &          \\
Lyubimkov \& Rachkovskaya (1995a,b)  &  8020   & 3.91  &    6.3  &    +0.27  &  7150 & 4.07  &  2.1  & $-0.33$  \\
Kunzli \& North (1998)               & (7581)  &(3.76) &   (3.2) &   (+0.15)$^{\dagger}$ &       &       &       &          \\
Hui-Bon-Hoa (2000)                   &  8240   & 4.00  &    3.0  &    +0.66  &  7610 & 4.10  &  2.5  & $-0.19$  \\
Khaliullin et al. (2001)              &  7570   & 3.89  &         &    +0.31$^{\dagger}$  &  6980 & 4.21  &       & $-0.24^{\dagger}$  \\
Jeong et al. (2017)                  &  7920   & 3.80  &    4.6  &    +0.20  &  7210 & 4.16  &  3.2  & $-0.34$  \\
Takeda et al. (2019) (Paper~I)       &  7570   & 3.90  &    3.8  &    +0.11  &  6980 & 4.21  &  3.1  & $-0.41$  \\
\hline
This study                           &  7990   & 3.90  &    3.5  &    +0.37  &  7570 & 4.20  &  2.6  & $-0.01$  \\                 
\hline
\end{tabular}
\end{center}
\scriptsize
The $T_{\rm eff}$ (in K), $\log g$ (in dex; $g$ is in unit of cm~s$^{-2}$), $v_{\rm t}$
(in km~s$^{-1}$) and [Fe/H] (Fe abundance relative to the Sun; in dex) adopted in the
past literature are given in column 2--5 (RR~Lyn A) and column 6--9 (RR~Lyn B). 
The parenthesized values for old studies (Smith 1971; Rachkovskaya 1974; Lyubimkov \& Rachkovskaya 1995b;
Kunzli \& North 1998) are those derived by regarding RR~Lyn as if a single star, 
which may be regarded as almost corresponding to the brighter A.
Regarding the projected rotational velocity ($v_{\rm e}\sin i$), Takeda et al. (2019) derived 
15.2 km~s$^{-1}$ (A) and 11.6~km~s$^{-1}$ (B) based on the spectrum-fitting analysis applied to 
the O~{\sc i} 7771--5 feature.\\
$^{\#}$Excitation temperature determined from Fe~{\sc i} lines.\\
$^{*}$Absolute abundance (in the usual normalization of H = 12) is given here, because the solar Fe abundance 
relevant to such an old analysis (based on unreliable $gf$ values as viewed from the present-day standard) 
is difficult to assign.\\
$^{\dagger}$Metallicity [M/H] derived from colors.\\  
\end{minipage}
\end{table*}

\section{Observational Data}

\subsection{Spectrum disentangling}

The basic observational materials are the high-dispersion spectra of RR~Lyn
(resolving power is $R \simeq 45000$) covering wide wavelength ranges ($\sim$~3800--9200~\AA), 
which were obtained on 2010 December 14, 15, 16, 18, and 20 by using BOES 
(Bohyunsan Observatory Echelle Spectrograph) attached to the 1.8 m reflector 
at Bohyunsan Optical Astronomy Observatory (cf. Sect.~2.1 and Table~2 of Paper~I for 
more details of these data).

Then, the separated spectra of A and B were numerically obtained by applying
the spectrum-disentangling technique to a set of nine double-line (A+B) 
spectra at 9 different phases,\footnote{The first spectrum on Dec. 14 coded 
``rrlyn\_20101214A'' (cf. Table~2 in Paper~I) was not used because it 
corresponds to the eclipse phase.}
 where the public-open software CRES\footnote{
http://sail.zpf.fer.hr/cres/} written by Dr. S. Iliji\'{c} was employed with 
the same procedure as detailed in Sect.~2.2 of Paper~I. 

Likewise, the same preconditioning processes as done in Paper~II were applied 
beforehand to the original spectra: (1) Continuum normalizations of all 
original spectra were done by regarding the wings of very strong lines (e.g., 
H lines) as if being the pseudo-continuum level, so that these features 
may be eventually wiped out in the final spectra. (2) Telluric features 
were removed in advance by dividing the spectrum by that of a rapid 
rotator or by erasing them interactively by hand on the screen.

In consequence, disentangled spectra of RR~Lyn A and B in 69 spectral regions 
(segments of typically several tens \AA\ to $\sim 100$~\AA\ wide; partially 
overlapped with each other) were obtained, which cover from $\sim 3900$~\AA\ 
to $\sim 9200$~\AA\ with a step of $\sim$~0.05~\AA.
The overall trends of these spectra are displayed in Fig.~1a (A) 
and 1b (B), and how the signal-to-noise ratio depend upon the wavelength 
is also shown in Fig.~1c (A) and 1d (B). We can see that (i) S/N for A 
($\lesssim$~500--800) is appreciably  higher than that for B ($\lesssim$~200) 
and (ii) the maximum S/N is attained around $\sim 6000$~\AA.    
All these spectra used for the analysis are included in the online 
materials (``specdata\_A.txt'' and ``specdata\_B.txt''). 

\begin{figure}[H] 
\begin{minipage}{85mm}
\begin{center}
   \includegraphics[width=8.5cm]{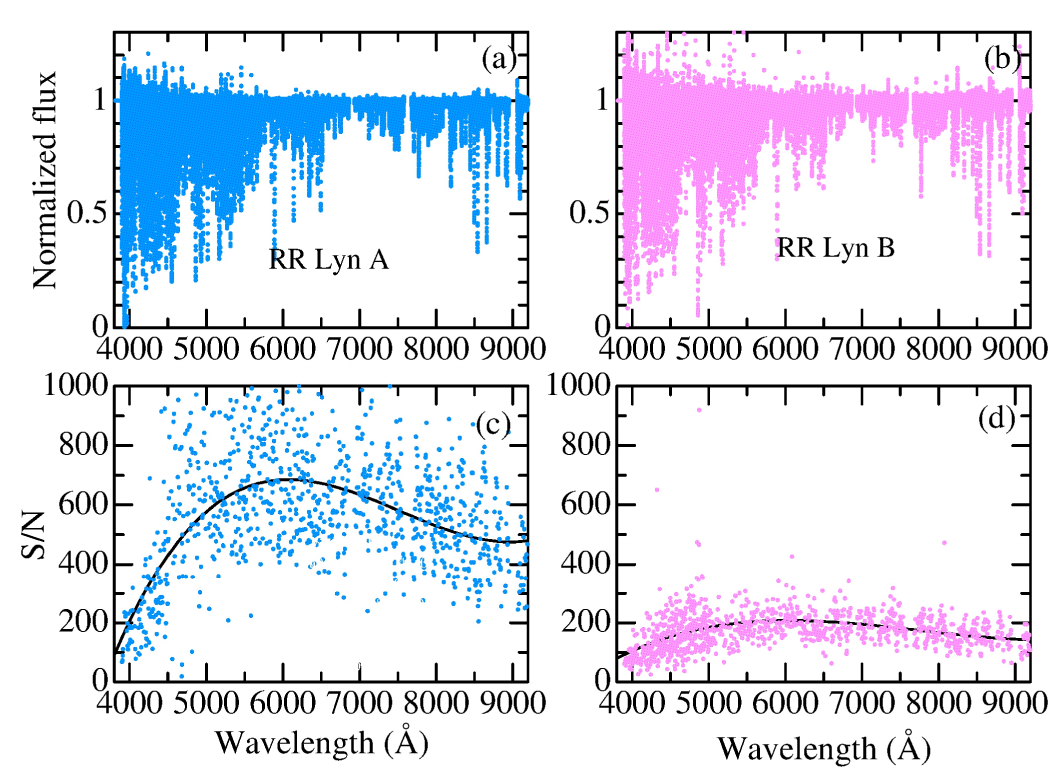}
\end{center}
\caption{The whole disentangled spectra of RR~Lyn A and B used in this 
study are plotted against wavelength in the upper panels (a) and (b).
The runs of signal-to-noise ratios of these spectra (directly estimated 
in line-free regions) are shown in the lower panels (c) and (d), 
where the fitted 3rd-order polynomials (describing the global trends) 
are also depicted by solid lines. 
} 
   \label{Fig1}
\end{minipage}
\end{figure}

\subsection{Line identification and equivalent widths}

Based on the disentangled spectra, lines usable for abundance determinations
were identified and their equivalent widths were measured.
The identification was done by carefully comparing the observed and theoretically 
synthesized spectra with each other. Here, lines to be measured were restricted 
to only those of single component (i.e., multiplet lines with fine structures 
such as Mg~{\sc ii} 4481 were discarded), and those being seriously affected 
by blending with other neighborhood lines were avoided.

The equivalent width ($W_{\lambda}$) of each line was evaluated by applying 
the Gaussian fitting to its profile. Fig.~2 demonstrates how the equivalent widths 
were actually measured for selected 24 lines of different element species.
All the data ($W_{\lambda}$ along with the atomic data taken from the VALD database;
cf. Ryabchikova et al. 2015) of finally identified 1364/1382 lines for A/B 
are also presented as the online material (``ewlines\_A.txt'' and 
``ewlines\_B.txt''). 

\begin{figure}[H]
\begin{minipage}{85mm}
\begin{center}
  \includegraphics[width=8.5cm]{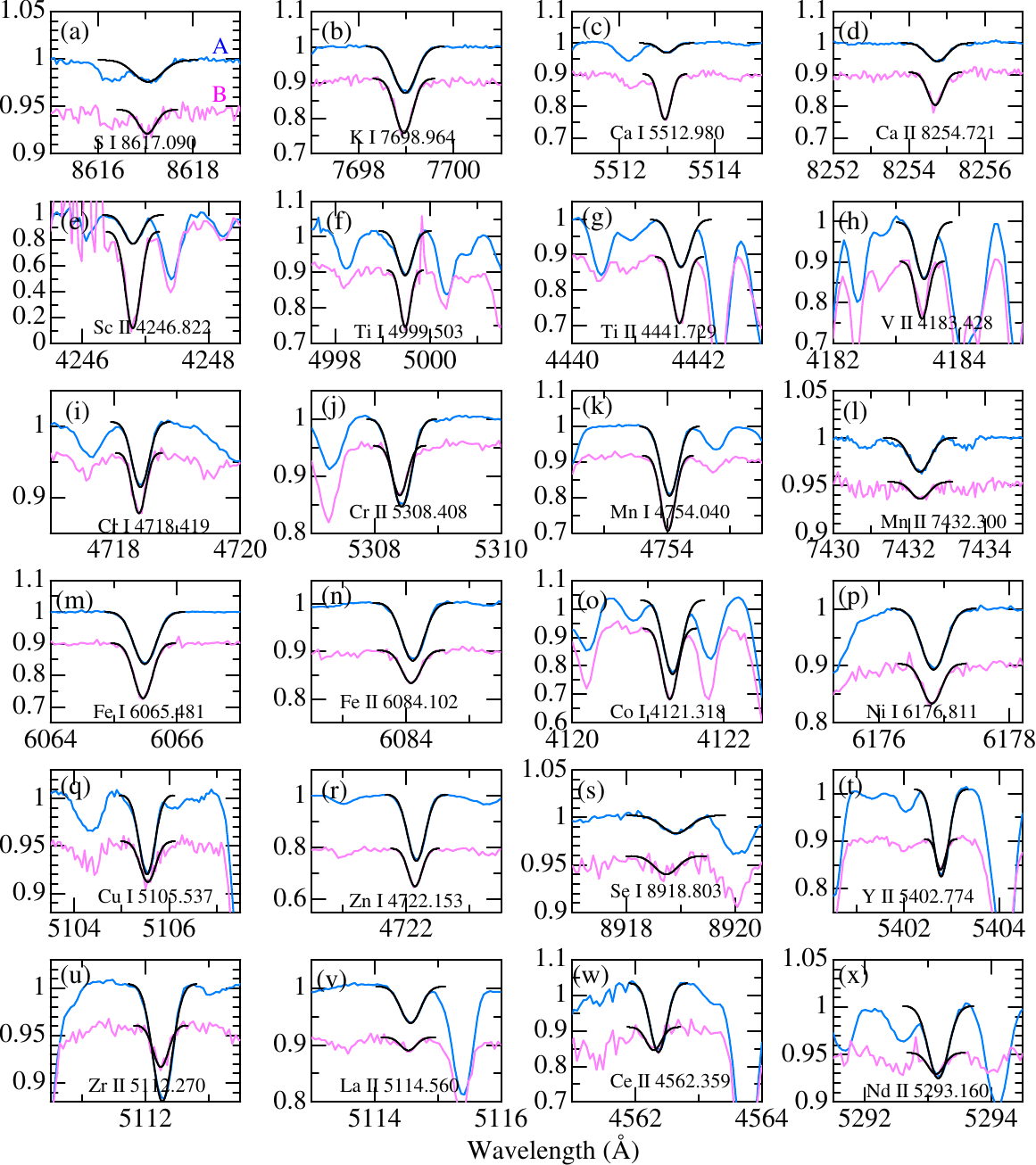}
\end{center}
   \caption{
In order to demonstrate how the equivalent widths were measured 
by the Gaussian fitting, selected cases of 24 lines (as indicated in each panel) 
are shown here. The observed spectra are drawn in color (blue for A, pink for B), 
while the fitted curve is depicted by a black line. The scale in the ordinate 
is for A, since the spectra for B are shifted downwards 
by appropriate amounts (0.05 or 0.1 or 0.2).
} 
   \label{Fig2}
\end{minipage}
\end{figure}

\section{Atmospheric Parameters}

\subsection{Basic principles}

One of the objectives of this study is to establish $T_{\rm eff}$ and $v_{\rm t}$ 
for both RR~Lyn A and B by making use of many Fe lines, as done in Paper~II.
As well known, $T_{\rm eff}$ can be determined by the requirement that the 
abundances ($A_{i}$) derived from the equivalent widths ($W_{i}$) of each line $i$ 
($i = 1, 2, \ldots, N$; where $N$ is the number of lines) do not systematically 
depend upon the lower excitation potential ($\chi_{\rm low}$) [excitation equilibrium], 
while $v_{\rm t}$ is determinable by demanding that $A_{i}$'s do not show any 
systematic trend irrespective of the equivalent widths ($W_{i}$) or reduced 
equivalent widths ($\equiv 10^{6} W_{i}/\lambda_{i}$) [curve-of-growth matching].
The desired solution ($T_{\rm eff}^{*}, v_{\rm t}^{*}$) simultaneously 
satisfying these two conditions can be obtained by finding the minimum 
of $\sigma(T_{\rm eff}, v_{\rm t})$, where $\sigma$ is the standard 
deviation of the abundances around the mean ($\langle A \rangle$) 
calculated for various combinations of $(T_{\rm eff}, v_{\rm t})$.  

\subsection{Line selection}

In the first place, some preparatory analysis may be in order, in order to define the 
set of lines to be used for the analysis by rejecting unsuitable lines.
For this purpose, Fe abundances ($A$) were calculated from the $W_{\lambda}$ values of 
all Fe lines measured in Sect.~2.2 by using the ($T_{\rm eff}$, $\log g$, $v_{\rm t}$)
values adopted in Paper~I [(7570, 3.90, 3.8) for A and (6980, 4.21, 3.1) for B]
as the starting trial parameters.
The resulting $A$ values are plotted against $10^{6} W_{\lambda}/\lambda$ and 
$\chi_{\rm low}$ in Fig.~3. An inspection of Fig.~3a,b (A) and Fig.~3e,f (B) shows that
a rather abrupt upturn of $A$ is seen at $\log (10^{6} W_{\lambda}/\lambda) \gtrsim 0.5$. 
Although the cause of this phenomenon is not clear (depth-dependence of $v_{\rm t}$?), 
it is certain that $W_{\lambda}$-independence of $A$ can not be accomplished as long as 
these stronger lines are included. 
Therefore, lines satisfying the condition $10^{6} W_{\lambda}/\lambda > 30$ 
(corresponding to $W_{\lambda} > 150$~m\AA\ at $\lambda = 5000$\AA) were discarded from the outset.  
Besides, outlier lines showing appreciable deviations were removed. The plots corresponding 
to those rejected lines are highlighted in light-green in Fig.~3.
After the elimination process, the numbers of Fe~{\sc i} and Fe~{\sc ii} lines ($N_{1}$, $N_{2}$) 
to be employed for the analysis in Sect.~3.3 are (522, 131) for A and (604, 78) for B.

\begin{figure}[H]
\begin{minipage}{85mm}
\begin{center}
  \includegraphics[width=8.5cm]{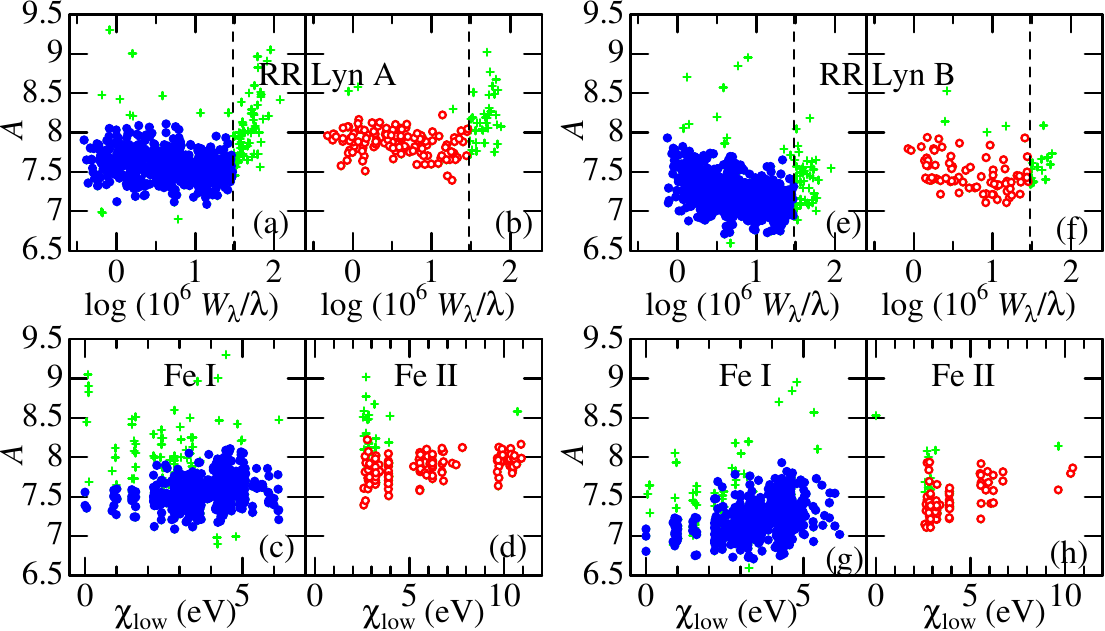}
\end{center}
   \caption{
Results of the preparatory analysis for selecting suitable Fe lines 
to be used for spectroscopic ($T_{\rm eff}$, $v_{\rm t}$) determination.
Plotted against $10^{6}W_{\lambda}/\lambda$ (reduced equivalent width)
and $\chi_{\rm low}$ (lower excitation potential) are the Fe abundances 
($A$) derived from the Fe~{\sc i} and Fe~{\sc ii} lines with the 
($T_{\rm eff}$, $v_{\rm t}$) values adopted in Paper~I [(7570~K, 3.8~km~s$^{-1}$) 
for A and (6980~K, 3.1~km~s$^{-1}$) for B]. The data corresponding to the 
finally adopted lines are depicted in blue filled circles (Fe~{\sc i}) and
red open circles (Fe~{\sc ii}), while those for the rejected lines
(lines showing outlier abundances or strong lines of $10^{6}W_{\lambda}/\lambda > 30$
(this critical limit is indicated by vertical dashed lines in the relevant panels)
are highlighted in light-green crosses. The four panels (a)--(d) on the left-hand side  
are for A and those of (e)--(h) on the right-hand side are for B.
} 
   \label{Fig3}
\end{minipage}
\end{figure}

\subsection{Solutions of $T_{\rm eff}$ and $v_{\rm t}$}

We are now ready to carry out an optimization analysis to obtain the desired solutions 
of ($T_{\rm eff}$, $v_{\rm t}$).   
For each species (Fe~{\sc i} or Fe~{\sc ii}) and each star (A or B), 
Fe abundances ($A_{i}, i=1, 2, \cdots, N$) 
were calculated from the equivalent widths ($W_{i}, i=1, 2, \cdots, N$) 
for an extensive grid of 1066 (=41$\times$26) 
cases resulting from combinations of 41 $T_{\rm eff}$ (from 6500 to 8500~K 
with a step of 50~K) and 26 $v_{\rm t}$ (from 1.0 to 6.0~km~s$^{-1}$ with 
a step of 0.2~km~$^{-1}$), where the necessary model atmospheres 
(solar metallicity models with $\log g = 3.9$ and $\log g =4.2$ for A and B; 
cf. Sect. 3.4) were generated by interpolating Kurucz's (1993) grid of ATLAS9 models. 
Then, $\langle A \rangle$ and $\sigma$ are calculated from the resulting set of 
($A_{i}, i=1, 2, \cdots, N$) for each of the 1066 combinations of ($T_{\rm eff}, v_{\rm t}$),
while a few outlier data (judged by Chauvenet's criterion) were discarded
in this process. 

The contour maps of such obtained $\sigma_{1{\rm A}}$, $\sigma_{2{\rm A}}$,
$\sigma_{1{\rm B}}$, and $\sigma_{2{\rm B}}$ on the $T_{\rm eff}$--$v_{\rm t}$
plane are depicted in Fig.~4a, 4b, 4c, and 4d, respectively.
The solutions of ($T_{\rm eff}^{*}$, $v_{\rm t}^{*}$) at the minimum of 
$\sigma$ (denoted by crosses in Fig.~4) for each case are summarized in Table~2.

The resulting ($T_{\rm eff}^{*}/v_{\rm t}^{*}/\langle A \rangle$) 
derived from Fe~{\sc i} and Fe~{\sc ii} lines are read from Table~2 as
(7990/3.5/7.85)$_{1{\rm A}}$ and (7920/4.0/7.88)$_{2{\rm A}}$ for A, while 
(7570/2.6/7.54)$_{1{\rm B}}$ and (7940/3.2/7.48)$_{2{\rm B}}$ for B,
where the values are appropriately rounded.  

\begin{figure}[H]
\begin{minipage}{85mm}
\begin{center}
  \includegraphics[width=8.5cm]{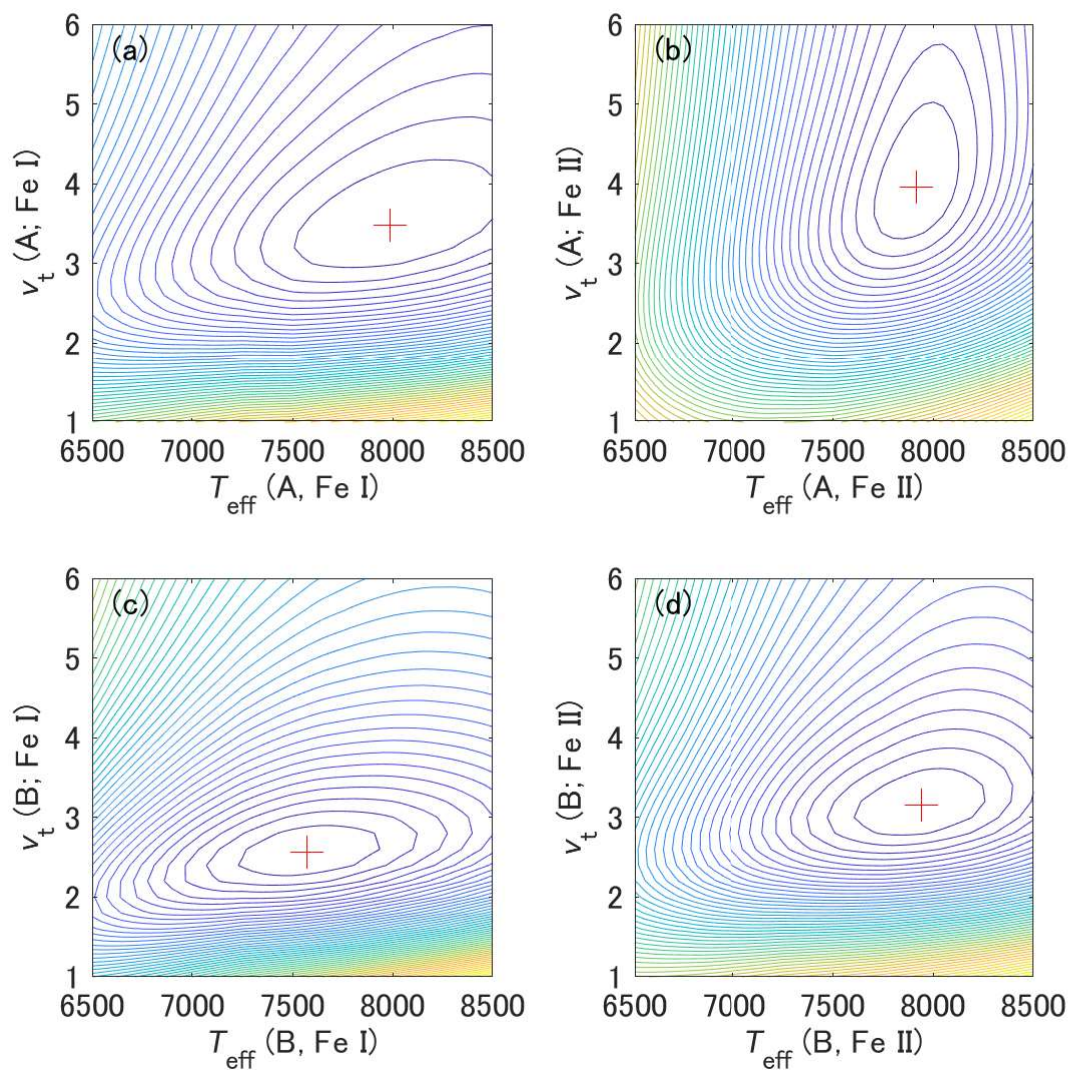}
\end{center}
   \caption{
Graphical display of the contours of $\sigma(T_{\rm eff}, v_{\rm t})$, 
where the position of $(T_{\rm eff}^{*}, v_{\rm t}^{*})$ 
corresponding the minimum of $\sigma$ is indicated by a cross.
The upper panels (a, b) are for RR~Lyn A, while the lower ones
(c, d) are for RR~Lyn B. The left panels (a, c) and the right ones (b, d)
are for Fe~{\sc i} and Fe~{\sc ii}, respectively. 
} 
   \label{Fig4}
\end{minipage}
\end{figure}

\subsection{Adopted parameters}

Two kinds of ($T_{\rm eff}^{*}, v_{\rm t}^{*}$) solutions were 
derived in Section~3.3 from Fe~{\sc i} and Fe~{\sc ii} lines. 
The next task is to decide which of these two solutions we should adopt,
since both are appreciably discrepant in some cases  
(e.g., $T_{\rm eff}$ for the case of B).

In contrast to the case of AR~Aur in Paper~II (where the results from Fe~{\sc ii}
were eventually adopted), we consider that the solutions derived from  
Fe~{\sc i} lines are more reliable than those from Fe~{\sc ii} lines for the 
present case of RR~Lyn.
This is because the number of Fe~{\sc ii} lines is distinctly smaller in comparison 
with Fe~{\sc i} lines (especially for B, where $N_{2}$ is only $\sim 1/8 N_{1}$)
and the scatter of $A_{2}$ is apparently too large to yield a convincing result
as can be seen in Fig.~3f and 3h.
Accordingly, the ($T_{\rm eff}^{*}, v_{\rm t}^{*}$) results derived from 
Fe~{\sc i} lines are adopted; i.e., (7990, 3.5) for A, and (7570, 2.6) for B.

The values of surface gravity are precisely determined for this eclipsing binary 
system; for example, Khaliullin e al. (2001) derived $3.894(\pm 0.019)$ (A) 
and $4.214(\pm 0.025)$ (B). Accordingly, $\log g = 3.90$ (A) and 4.20 (B) are adopted
in this study (essentially the same as in Paper~I) with an uncertainty of 
$\lesssim 0.05$~dex, which is practically sufficient because abundances are  less
sensitive to this parameter.
As to the model metallicity, solar composition models are adopted (as done in 
Paper~I and Paper~II), which is sufficient because atmospheric structures of 
early-type stars do not depend much upon the metallicity.  

The final parameters of RR~Lyn A and B used 
for abundance determinations are summarized in Table~3, where 
probable uncertainties are also given.  
The Fe abundances ($A_{i}$) derived from each of Fe~{\sc i} and 
Fe~{\sc ii} lines corresponding to the adopted Fe~{\sc i}-based 
$T_{\rm eff}^{*}/v_{\rm t}^{*}$ solutions are plotted against 
($10^{6}W_{i}/\lambda_{i}$) and $\chi_{\rm low}$ and the relevant  
empirical curves of growth are depicted in the left (Fe~{\sc i}) and 
center (Fe~{\sc ii}) columns of Fig.~5 (A) and Fig.~6 (B),
where the Fe~{\sc ii} results for the (unadopted) Fe~{\sc ii}-based 
parameters are also shown in the right columns for comparison.
It can be seen from these figures that the required condition 
(no systematic dependence in $A_{i}$ upon $W_{i}$ and $\chi_{\rm low}$)
is almost satisfied.

\setcounter{table}{1}
\begin{table}[h]
\begin{minipage}{80mm}
\begin{center}
\caption{($T_{\rm eff}$, $v_{\rm t}$) solutions at the minimum of $\sigma$.}
\small
\begin{tabular}{cccccc}
\hline\hline                 
Lines & $T_{\rm eff}^{*}$ & $v_{\rm t}^{*}$  &  $\sigma^{*}$ & $\langle A \rangle$ & Fig. \\
      &  (K)          & (km~s$^{-1}$)  &  (dex)     & (dex)    &  \\
\hline
\multicolumn{6}{c}{[RR Lyn A]} \\
 Fe~{\sc i}    & 7989  &   3.48  &   0.177 &  7.851 & 4a \\ 
               & (175)  &  (0.26) &         &       &    \\
 Fe~{\sc ii}   & 7916  &   3.96  &   0.128 &  7.881 & 4b \\
               & (94)  &  (0.38) &         &        &    \\
\hline
\multicolumn{6}{c}{[RR Lyn B]} \\
 Fe~{\sc i}    & 7575  &    2.56  &   0.193 &  7.537 & 4c \\ 
               & (199)  &  (0.16) &         &        &    \\
 Fe~{\sc ii}   & 7943  &   3.15  &   0.142 &  7.475  & 4d \\
               & (182)  &  (0.26) &         &        6    \\
\hline
\end{tabular}
\end{center}
\scriptsize
Columns 2 and 3 give the values of $T_{\rm eff}$ and $v_{\rm t}$,  
at which Fe abundance dispersion is minimized. The corresponding 
$\sigma$ and the mean Fe abundance are presented in columns 4
and 5, respectively.  See the figures indicated in column 6 
for the relevant $\sigma (T_{\rm eff}, v_{\rm t})$ contours.
The parenthesized values are the probable uncertainties involved
in $T_{\rm eff}^{*}$ and $v_{\rm t}^{*}$, which were estimated from 
$\sigma^{*}$ by random simulations as described in Sect.~3.3 of 
Takeda (2024). 
\end{minipage}
\end{table}

\begin{figure}[H] 
\begin{minipage}{85mm}
\begin{center}
   \includegraphics[width=8.5cm]{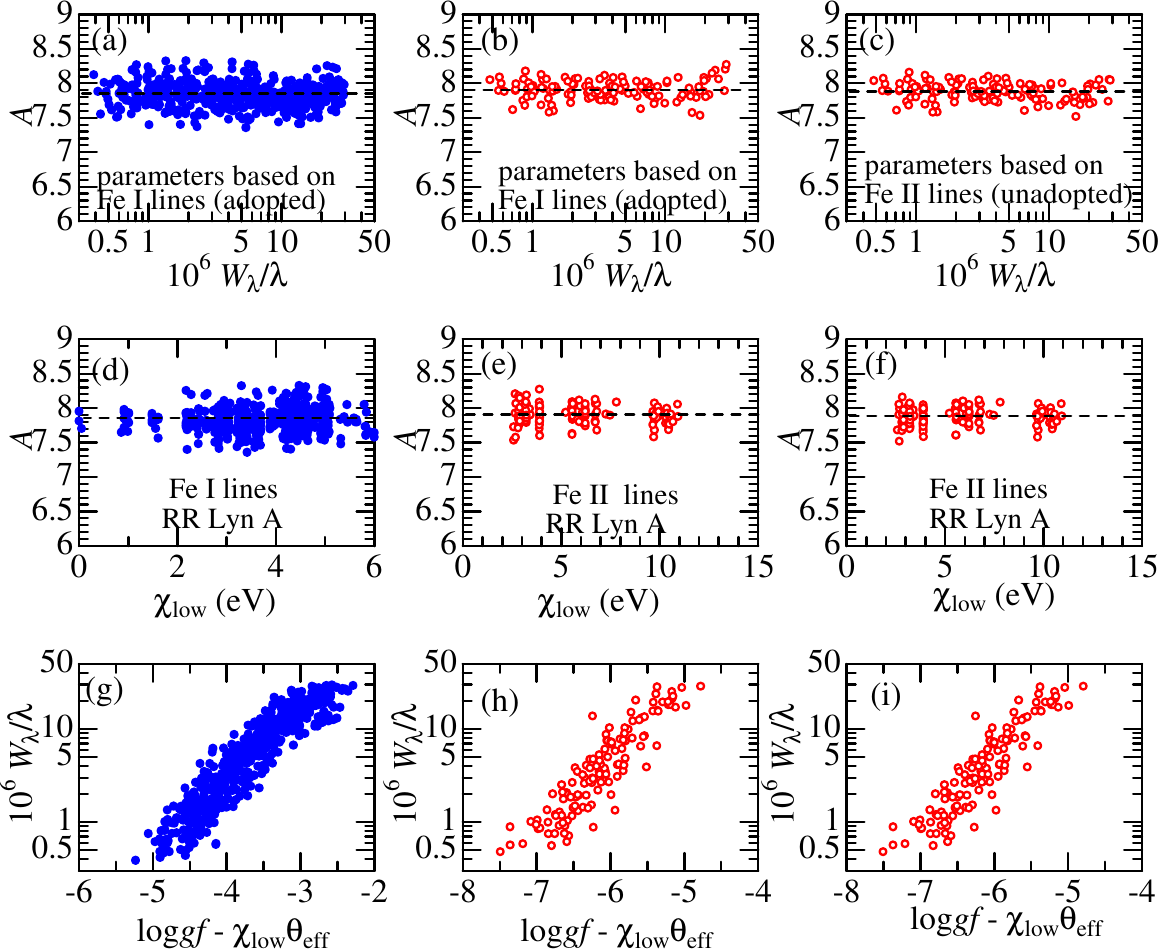}
\end{center}
\caption{
Fe abundances ($A_{i}$) of RR~Lyn~A corresponding to two kinds of ($T_{\rm eff}$, $v_{\rm t}$) 
solutions are plotted against the reduced equivalent width ($10^{6}W_{\lambda}/\lambda$; 
top panels) or lower excitation potential ($\chi_{\rm low}$; middle panels). 
Left and center panels correspond to the $\sigma_{1}$-minimum solutions based on 
Fe~{\sc i} lines (finally adopted), while the right panels correspond to $\sigma_{2}$-minimum 
solutions based on Fe~{\sc ii} lines (which were not adopted after all).
The mean abundance ($\langle A \rangle$) is also indicated by the 
horizontal dashed line. In the bottom panels are shown the corresponding empirical 
curves of growth, where $\log gf - \chi_{\rm low}(5040/T_{\rm eff})$ is taken 
as the abscissa. The results for Fe~{\sc i} and Fe~{\sc ii} lines are distinguished
by blue filled circles and red open circles, respectively.
} 
   \label{Fig5}
\end{minipage}
\end{figure}

\begin{figure}[H] 
\begin{minipage}{85mm}
\begin{center}
   \includegraphics[width=8.5cm]{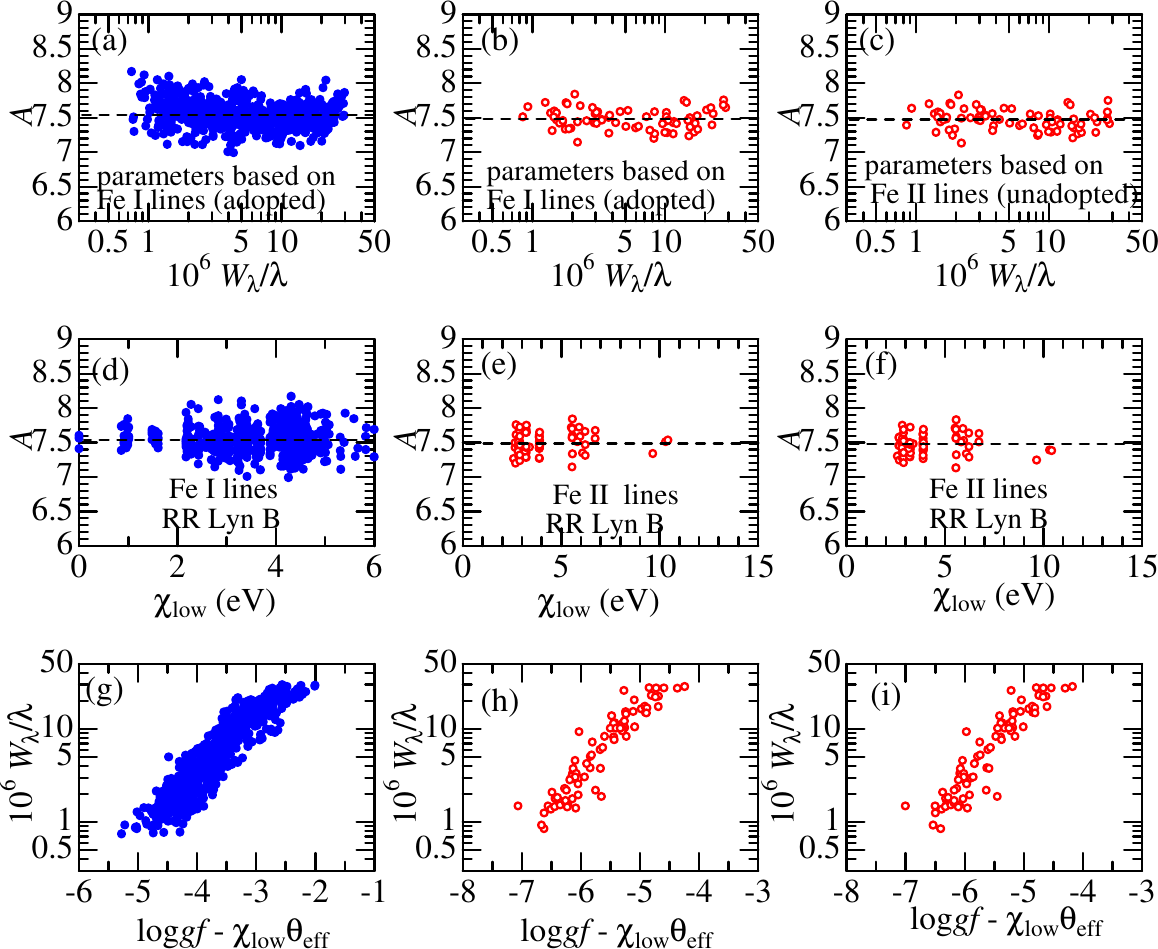}
\end{center}
\caption{
The results of Fe abundances ($A_{i}$) for RR~Lyn~B are presented. The left and 
center panels correspond to $\sigma_{1}$-minimum solutions of ($T_{\rm eff}$, $v_{\rm t}$) 
based on Fe~{\sc i} lines (finally adopted), while the right panels are for the 
$\sigma_{2}$-minimum solutions based on Fe~{\sc ii} lines (which were not adopted). 
Otherwise, the same as in Fig.~5.
} 
   \label{Fig6}
\end{minipage}
\end{figure}

\setcounter{table}{2}
\begin{table}[h]
\begin{minipage}{80mm}
\begin{center}
\caption{Adopted atmospheric parameters.}
\small
\begin{tabular}{cccc}
\hline\hline                 
Star & $T_{\rm eff}$ & $v_{\rm t}$ & $\log g^{\#}$ \\
     &  (K)          & (km~s$^{-1}$ & (dex)   \\
\hline
A    & 7990        &        3.5   &    3.90      \\ 
     & ($\pm 200$)  &  ($\pm 0.3$) & ($\pm 0.05$) \\
\hline
B    & 7570        &        2.6   &    4.20      \\
     & ($\pm 200$)  &  ($\pm 0.2$) & ($\pm 0.05$) \\
\hline
\end{tabular}
\end{center}
\scriptsize
Given here are the model atmosphere parameters (based on the solutions 
obtained from Fe~{\sc i} lines; cf. Table~2) finally adopted for 
deriving the chemical abundances of RR~Lyn A and B.
Parenthesized values are the roughly assigned uncertainties, which 
are based on the simulated random errors for $T_{\rm eff}$ and $\log g$ 
(cf. Table~2) and from the $\log g$ errors of a few hundredths dex 
given by Khaliullin et al. (2001).
\\
$^{\#}g$ is in cm~s$^{-2}$.
\end{minipage}
\end{table}

\section{Abundance determination}

\subsection{Direct and synthetic equivalent widths}

Based on the model atmospheres with the atmospheric parameters 
established in Section~3.4 (Table~3), elemental abundances of A and B are 
determinable from the equivalent widths ($W_{\lambda}$). 

However, since identification and $W_{\lambda}$ measurement by Gaussian 
fitting done in Section~2.2 was restricted to single-component lines without 
any serious blending with other lines, multi-component lines (e.g., 
Li~{\sc i} 6708 or Ba~{\sc ii} 5854) or important lines appreciably blended 
with other features could not be included. 

Therefore, an alternative synthetic spectrum-fitting approach 
was additionally applied to selected line features to evaluate the 
relevant $W_{\lambda}$ (corresponding to the target species) inversely 
from the resulting abundance. Regarding the details of this alternative 
$W_{\lambda}$-determination approach, Sect.~4 of Takeda et al. (2018) 
may be consulted. This synthetic fitting analysis was conducted for 
36 regions in total. Fig.~7 illustrates the demonstrative examples of 
selected 12 regions. 

\begin{figure}[H]
\begin{minipage}{85mm}
\begin{center}
  \includegraphics[width=8.5cm]{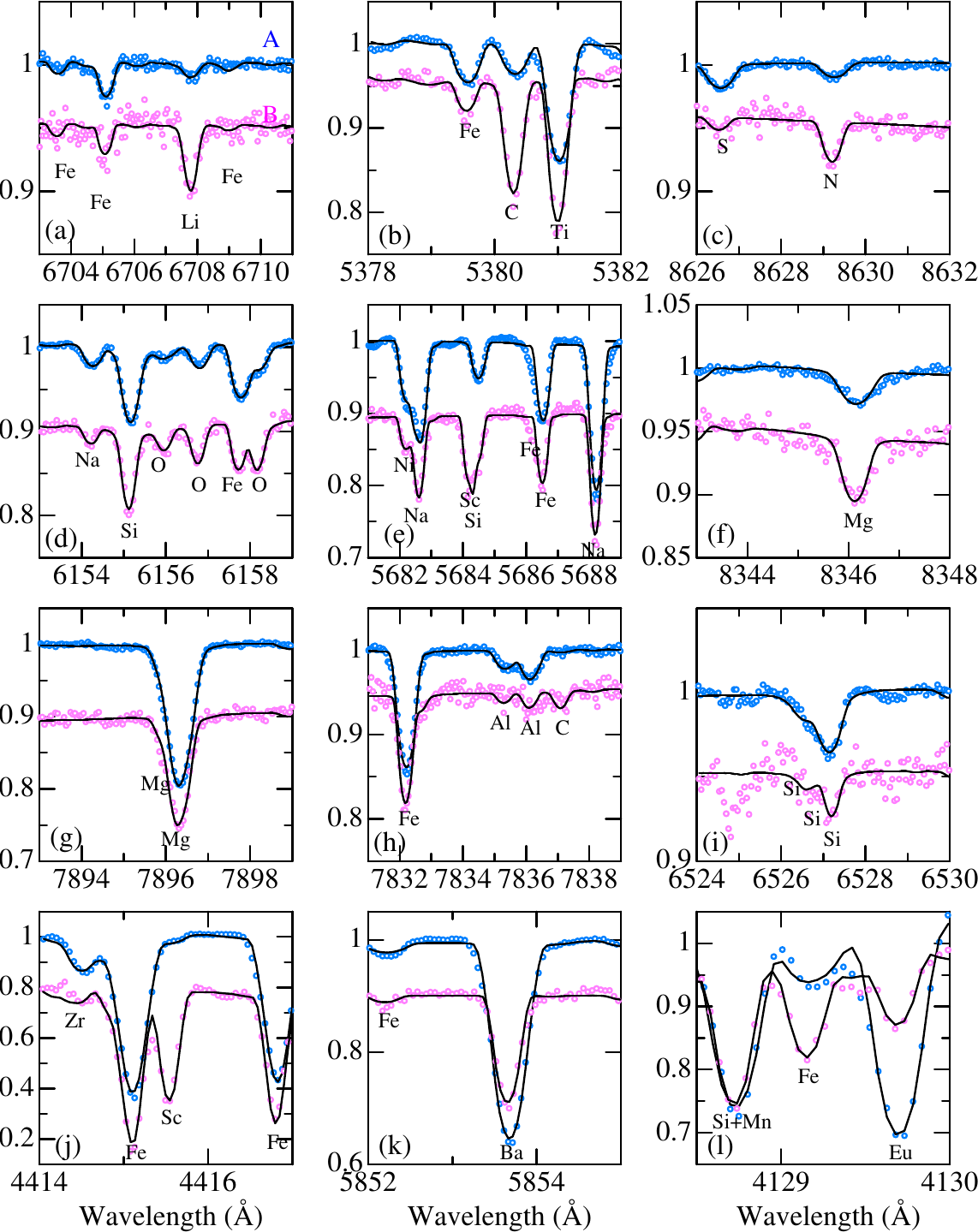}
\end{center}
   \caption{
The accomplished fit of the synthetic spectrum-fitting analysis 
carried out for the purpose of inversely evaluating the equivalent widths.
Shown here are the demonstrative results for 12 selected regions, each of 
which include the following target lines: (a) Li~{\sc i} 6708, (b) C~{\sc i} 5380,
(c) N~{\sc i} 8629, (d) O~{\sc i} 6156/6157/6158, (e) Na~{\sc i} 5682/5688,
(f) Mg~{\sc i} 8346, (g) Mg~{\sc ii} 7896, 
(h) Al~{\sc i} 7835/7836, (i) Si~{\sc i} 6527,
(j) Sc~{\sc ii} 4415, (k) Ba~{\sc ii} 5854, and (l) Eu~{\sc ii} 4130. 
The observed and theoretical spectra are depicted by symbols 
(blue for A, pink for B) and black lines, respectively. 
As in Fig.~2,  the scale marked in the ordinate is for A, since 
the spectra for B are shifted downwards by appropriate amounts 
(0.05 or 0.1 or 0.2). 
} 
   \label{Fig7}
\end{minipage}
\end{figure}

\subsection{Non-LTE calculations}

According to the policy of taking the non-LTE effect into consideration 
wherever possible, non-LTE abundances were derived for comparatively lighter 
elements ($Z \le 20$ and $Z = 30$), for which the author already has experiences 
of non-LTE calculations. For other elements ($Z \ge 21$ except for Zn) the abundances 
were derived under the assumption of LTE. The finally adopted (input) abundances 
in the calculations (and the related papers) are summarized in Table~4.

Regarding silicon (Si, $Z = 14$), although its non-LTE calculation was carried 
out by Takeda (2022) in his analysis of Si~{\sc ii} 6347/6371 doublet lines in 
A- and B-type stars, the atomic model used therein (34 Si~{\sc i} terms up 
to 58893~cm$^{-1}$ with 222 Si~{\sc i} radiative transitions) did not 
comprise sufficient number of high-excitation Si~{\sc i} levels, because
neutral Si~{\sc i} lines were outside of the subject in that paper.
Given the necessity of analyzing many Si~{\sc i} lines in this study, the 
atomic model for Si~{\sc i} was upgraded to 102 Si~{\sc i} terms (up to 
63945~cm$^{-1}$) with 1555 Si~{\sc i} radiative transitions.
Other details of the model atom (e.g., Si~{\sc ii} model, photoionization 
or collisional rates) are the same as described in Sect.~4 of Takeda (2022).

\setcounter{table}{3}
\begin{table}[h]
\begin{minipage}{80mm}
\begin{center}
\caption{Non-LTE calculations done in this study.}
\small
\begin{tabular}{c@{ }cc@{ }cl}
\hline\hline\noalign{\smallskip}
Elem. & $Z^{\#}$ & [X/H]$_{\rm A}^{*}$ & [X/H]$_{\rm B}^{*}$ & References$^{\dagger}$ \\
\hline\noalign{\smallskip}
 Li   &  3  &  $-0.4$  & $ 0.0$ & Takeda \& Kawanomoto (2005) \\
 C    &  6  &  $-0.8$  & $-0.3$ & Takeda (1992) \\
 N    &  7  &  $-1.1$  & $-0.6$ & Takeda (1992) \\
 O    &  8  &  $-0.9$  & $-0.4$ & Takeda (2003) \\
 Na   & 11  &  $+0.2$  & $-0.2$ & Takeda \& Takada-Hidai (1994) \\
 Mg   & 12  &  $-0.3$  & $-0.3$ & Takeda (2025) (Paper II)\\
 Al   & 13  &  $+0.4$  & $-0.2$ & Takeda (2023) \\
 Si   & 14  &  $+0.1$  & $-0.1$ & Takeda (2022)$^{\ddagger}$ \\
 S    & 16  &  $+0.2$  & $+0.1$ & Takeda et al. (2005) \\
 K    & 19  &  $-0.1$  & $-0.3$ & Takeda et al. (1996) \\
 Ca   & 20  &  $-0.3$  & $ 0.0$ & Takeda (2020) \\
 Zn   & 30  &  $+0.9$  & $+0.1$ & Takeda et al. (2005) \\
\hline
\end{tabular}
\end{center}
\scriptsize
$^{\#}$Atomic number.\\
$^{*}$Abundances (relative to the solar composition)
assigned in non-LTE calculations for each star (in dex), which were 
(iteratively) chosen so that they may be almost consistent with 
the final results of non-LTE abundances.\\
$^{\dagger}$These papers (and the references quoted therein) may be consulted
for more details about the calculations (e.g., adopted model atoms).\\
$^{\ddagger}$See also Sect.~4.2.\\
\end{minipage}
\end{table}

\subsection{Abundance results}

By inspecting the list of identified/measured lines prepared in Sect.~2.2 
(or the list of selected lines defined in Sect.~3.2 for Fe), 
we focus on the following 39 species of 30 elements:
Li~{\sc i}, C~{\sc i}, N~{\sc i}, O~{\sc i}, Na~{\sc i},
Mg~{\sc i}, Mg~{\sc ii}, Al~{\sc i}, Al~{\sc ii}, Si~{\sc i}, Si~{\sc ii}, 
S~{\sc i}, K~{\sc i}, Ca~{\sc i}, Ca~{\sc ii}, Sc~{\sc ii}, Ti~{\sc i}, 
Ti~{\sc ii}, V~{\sc i}, V~{\sc ii}, Cr~{\sc i}, Cr~{\sc ii}, 
Mn~{\sc i}, Mn~{\sc ii}, Fe~{\sc i}, Fe~{\sc ii}, Co~{\sc i}, Ni~{\sc  i}, 
Zn~{\sc i}, Cu~{\sc i}, Se~{\sc i}, Sr~{\sc ii}, Y~{\sc ii}, Zr~{\sc ii}, 
Ba~{\sc ii}, La~{\sc ii}, Ce~{\sc ii}, Nd~{\sc ii}, and Eu~{\sc ii}.
Here, the lines to be adopted are restricted to only those for which 
$W_{\lambda}$ values of ``both A and B'' are measured and available, 
because of an intention of line-by-line differential analysis.
 
Based on the equivalent widths of these selected lines 
(derived either by direct measurement or synthetic fitting),
the abundances of relevant elements were determined by using
Kurucz's (1993) WIDTH9 program, which was considerably modified
by the author (e.g., treatment of merged multi-component features, 
taking into account the non-LTE effect, etc.).
All atomic data were taken from VALD (Ryabchikova et al. 2015),
except for the Li~{\sc i} 6708 feature for which the recipe described in
Sect.~2.2 of Takeda \& Kawanomoto (2005) was followed (but neighboring molecular
lines were not included).  
The resulting detailed line-by-line abundances for A and B (with 
non-LTE corrections if available) and their line-averaged abundances, 
along with the corresponding $W_{\lambda}$ and atomic data of 
spectral lines, are presented in ``abundances.txt'' of the online material.

Further, in deriving the mean abundances, the data of strong lines 
satisfying $10^{6}W_{\lambda}/\lambda > 30$ were excluded for the same reason
as mentioned in Sect.~3.2 for the case of Fe lines in the determination of 
($T_{\rm eff}$, $v_{\rm t}$).
 
Table~5 presents the mean results averaged over the available lines 
for each species: $\langle$[X/H]$_{\rm A}\rangle$ or
$\langle$[X/H]$_{\rm B}\rangle$ (mean abundance relative to 
the Sun; i.e., line-average of $A_{i} - A_{\odot}$), and 
$\langle \Delta A^{\rm X}_{\rm A-B}\rangle$ (mean of line-by-line
difference between A and B; i.e., line-average of
$A_{i,{\rm A}} - A_{i,{\rm B}}$). The reference solar abundances 
($A_{\odot}$; given in the 4th column of Table~5)
were mostly taken from Anders \& Grevesse (1989)\footnote{These data may be 
somewhat outdated for some elements as compared to more recent compilations. 
For example, Asplund et al.'s (2009) solar CNO abundances (derived by taking 
into account the 3D effect) are by $\sim$~0.2--0.3~dex lower than the values 
adopted here. Note, however, that the solar CNO abundances derived from 
the classical 1D plane-parallel atmosphere by including the non-LTE effect 
(such as done in this study) are rather consistent with those of Anders \& 
Grevesse (1989) adopted here (cf. Takeda 1994).} in order to maintain consistency 
with the author's previous studies. The exceptions are 3.28 for Li (solar system 
abundance taken from Lodders 2003) and 7.50 for Fe (currently accepted value 
for the solar photosphere).   

The sensitivities of abundance results to changing the atmospheric parameters 
($T_{\rm eff}$, $v_{\rm t}$, and $\log g$) are also presented in Table~5,
where we can see from this table that the impact of $T_{\rm eff}$ is more 
significant than the other two. 
That is, abundance errors due to uncertainties in $T_{\rm eff}$ 
($\sim \pm 200$~K; cf. Table~3) are $\lesssim$~0.1--0.2~dex, while those due to
$v_{\rm t}$ ($\pm 0.3$~km~s$^{-1}$) as well as $\log g$ ($\pm 0.05$~dex)
are only a few hundredths dex in most cases (note that the $\Delta$ 
values given in Table~5 correspond to perturbations of $T_{\rm eff}$ by 
300~K, $v_{\rm t}$ by 0.3~km~s$^{-1}$, and $\log g$ by 0.1~dex; i.e., 
somewhat larger compared to above-mentioned ambiguities 
regarding $T_{\rm eff}$ and $\log g$).

\setcounter{table}{4}
\begin{table*}[h]
\begin{minipage}{160mm}
\begin{center}
\caption{Results of elemental abundances for RR~Lyn A and B.}
\small
\begin{tabular}{cccccc@{ }cc@{ }cc@{ }cc@{ }cc@{ }cc@{ }cl
}\hline\hline
$Z$ & Species & $T_{\rm c}$ & $A_{\odot}$ & $N$ & $\langle$[X]$_{\rm A}\rangle$  & 
$N_{\rm A}$ & $\langle$[X]$_{\rm B}\rangle$  & $N_{\rm B}$ &  
$\langle \Delta X_{\rm A-B}\rangle$ & $N_{\rm A-B}$ &
$\Delta^{T+}_{\rm A}$ & $\Delta^{T+}_{\rm B}$ &
$\Delta^{v+}_{\rm A}$ & $\Delta^{v+}_{\rm B}$ &
$\Delta^{g+}_{\rm A}$ & $\Delta^{g+}_{\rm B}$ \\
(1) & (2) & (3) & (4) & (5) & (6) & (7) & (8) & (9) & (10) &
(11) & (12) & (13) & (14) & (15) & (16) & (17) \\ 
\hline
\multicolumn{17}{c}{(non-LTE analysis)}\\
 3 &  Li~{\sc i}  & 1142 &  3.28 &  1 &  $-$0.43& (  1)& $-$0.07& (  1)& $-$0.36& (  1)&    +16& +11&  +00& +00&  $-$01& +01  \\
 6 &  C~{\sc i}   &   40 &  8.56 & 25 &  $-$0.82& ( 23)& $-$0.33& ( 20)& $-$0.52& ( 20)&    +03& $-$03&  $-$01& $-$01&  +01& +02  \\
 7 &  N~{\sc i}   &  123 &  8.05 &  7 &  $-$1.06& (  6)& $-$0.56& (  6)& $-$0.56& (  7)&    $-$04& $-$08&  $-$00& $-$00&  +02& +03  \\
 8 &  O~{\sc i}   &  180 &  8.93 &  9 &  $-$0.98& (  7)& $-$0.41& (  8)& $-$0.53& (  7)&    $-$05& $-$09&  $-$01& $-$02&  +02& +03  \\
11 &  Na~{\sc i}  &  958 &  6.33 &  8 &  +0.24& (  5)& $-$0.19& (  6)& +0.43& (  4)&    +11& +07&  $-$02& $-$02&  $-$01& +00  \\
12 &  Mg~{\sc i}  & 1336 &  7.58 & 18 &  $-$0.31& ( 13)& $-$0.29& ( 13)& $-$0.04& ( 12)&    +12& +07&  $-$01& $-$01&  $-$01& +00  \\
12 &  Mg~{\sc ii} & 1336 &  7.58 &  6 &  $-$0.25& (  4)& $-$0.25& (  4)& +0.01& (  5)&    $-$03& $-$08&  $-$02& $-$01&  +01& +03  \\
13 &  Al~{\sc i}  & 1653 &  6.47 &  8 &  +0.31& (  6)& $-$0.16& (  6)& +0.47& (  6)&    +13& +06&  $-$00& $-$00&  $-$02& +01  \\
13 &  Al~{\sc ii} & 1653 &  6.47 &  1 &  +0.45& (  1)& +0.33& (  1)& +0.12& (  1)&    $-$05& $-$10&  $-$01& $-$00&  +02& +03  \\
14 &  Si~{\sc i}  & 1310 &  7.55 & 49 &  +0.13& ( 47)& $-$0.04& ( 48)& +0.20& ( 47)&    +11& +06&  $-$00& $-$01&  $-$01& +01  \\
14 &  Si~{\sc ii} & 1310 &  7.55 &  5 &  +0.17& (  4)& $-$0.09& (  5)& +0.25& (  4)&    $-$04& $-$09&  $-$03& $-$03&  +02& +04  \\
16 &  S~{\sc i}   &  664 &  7.21 & 13 &  +0.24& ( 11)& +0.12& ( 11)& +0.16& ( 12)&    +09& +02&  $-$01& $-$01&  +00& +02  \\
19 &  K~{\sc i}   & 1006 &  5.12 &  1 &  $-$0.08& (  1)& $-$0.33& (  1)& +0.24& (  1)&    +13& +09&  $-$01& $-$02&  $-$01& +00  \\
20 &  Ca~{\sc i}  & 1517 &  6.36 & 26 &  $-$0.42& ( 24)& +0.01& ( 21)& $-$0.41& ( 21)&    +15& +08&  $-$01& $-$05&  $-$01& +01  \\
20 &  Ca~{\sc ii} & 1517 &  6.36 &  5 &  $-$0.25& (  4)& +0.04& (  5)& $-$0.29& (  4)&    +02& $-$03&  $-$04& $-$05&  +02& +02  \\
30 &  Zn~{\sc i}  &  726 &  4.60 &  4 &  +0.77& (  4)& +0.10& (  4)& +0.67& (  4)&    +14& +08&  $-$03& $-$02&  $-$01& +01  \\
\hline
\multicolumn{17}{c}{(LTE analysis)}\\
21 &  Sc~{\sc ii} & 1659 &  3.10 &  7 &  $-$1.57& (  6)& +0.21& (  6)& $-$1.83& (  5)&    +10& +06&  $-$00& $-$08&  +03& +04  \\
22 &  Ti~{\sc i}  & 1582 &  4.99 &  4 &  $-$0.01& (  4)& +0.00& (  4)& $-$0.01& (  4)&    +16& +10&  $-$00& $-$01&  $-$01& +01  \\
22 &  Ti~{\sc ii} & 1582 &  4.99 & 60 &  $-$0.04& ( 38)& +0.04& ( 41)& $-$0.07& ( 38)&    +09& +05&  $-$02& $-$03&  +03& +04  \\
23 &  V~{\sc i}   & 1429 &  4.00 &  1 &  +0.07& (  1)& $-$0.22& (  1)& +0.29& (  1)&    +18& +13&  $-$00& $-$00&  $-$01& +01  \\
23 &  V~{\sc ii}  & 1429 &  4.00 &  6 &  +0.32& (  6)& +0.01& (  5)& +0.28& (  5)&    +08& +05&  $-$02& $-$02&  +03& +04  \\
24 &  Cr~{\sc i}  & 1296 &  5.67 & 21 &  +0.23& ( 19)& $-$0.08& ( 19)& +0.31& ( 19)&    +15& +10&  $-$01& $-$02&  $-$01& +01  \\
24 &  Cr~{\sc ii} & 1296 &  5.67 & 32 &  +0.43& ( 24)& +0.05& ( 29)& +0.40& ( 24)&    +05& +01&  $-$03& $-$03&  +03& +04  \\
25 &  Mn~{\sc i}  & 1158 &  5.39 & 11 &  +0.26& (  9)& $-$0.08& (  7)& +0.36& (  9)&    +15& +09&  $-$01& $-$02&  $-$01& +01  \\
25 &  Mn~{\sc ii} & 1158 &  5.39 &  1 &  +0.31& (  1)& $-$0.08& (  1)& +0.39& (  1)&    +06& +02&  $-$01& $-$00&  +02& +04  \\
26 &  Fe~{\sc i}  & 1134 &  7.50 &424 &  +0.34& (424)& +0.01& (424)& +0.33& (420)&    +14& +09&  $-$02& $-$02&  $-$01& +01  \\
26 &  Fe~{\sc ii} & 1134 &  7.50 & 56 &  +0.40& ( 56)& $-$0.03& ( 55)& +0.42& ( 52)&    +05& +01&  $-$03& $-$01&  +03& +04  \\
27 &  Co~{\sc i}  & 1352 &  4.92 &  5 &  +0.50& (  5)& +0.05& (  5)& +0.45& (  5)&    +16& +12&  $-$02& $-$02&  $-$01& +01  \\
28 &  Ni~{\sc i}  & 1353 &  6.25 & 80 &  +0.63& ( 77)& +0.05& ( 79)& +0.60& ( 74)&    +14& +09&  $-$01& $-$01&  $-$01& +01  \\
29 &  Cu~{\sc i}  & 1037 &  4.21 &  1 &  +0.71& (  1)& +0.07& (  1)& +0.65& (  1)&    +17& +12&  $-$01& $-$01&  $-$01& +01  \\
34 &  Se~{\sc i}  &  697 &  3.35 &  1 &  +0.51& (  1)& +0.27& (  1)& +0.24& (  1)&    +12& +05&  $-$00& $-$00&  $-$00& +02  \\
38 &  Sr~{\sc ii} & 1464 &  2.90 &  4 &  +0.99& (  1)& +0.37& (  2)& +0.52& (  1)&    +11& +04&  $-$05& $-$04&  +02& +04  \\
39 &  Y~{\sc ii}  & 1659 &  2.24 & 12 &  +1.28& (  7)& +0.45& ( 11)& +0.84& (  7)&    +11& +07&  $-$03& $-$03&  +02& +04  \\
40 &  Zr~{\sc ii} & 1741 &  2.60 &  8 &  +1.19& (  5)& +0.54& (  6)& +0.73& (  6)&    +09& +06&  $-$02& $-$01&  +03& +04  \\
56 &  Ba~{\sc ii} & 1455 &  2.13 &  4 &  +1.51& (  1)& +0.54& (  3)& +1.20& (  1)&    +19& +12&  $-$11& $-$11&  +01& +03  \\
57 &  La~{\sc ii} & 1578 &  1.22 &  5 &  +1.25& (  3)& +0.46& (  5)& +0.77& (  3)&    +16& +10&  $-$02& $-$02&  +02& +04  \\
58 &  Ce~{\sc ii} & 1478 &  1.55 &  4 &  +1.30& (  4)& +0.52& (  4)& +0.78& (  4)&    +13& +09&  $-$01& $-$00&  +02& +04  \\
60 &  Nd~{\sc ii} & 1602 &  1.50 &  2 &  +1.10& (  2)& +0.22& (  2)& +0.88& (  2)&    +17& +11&  $-$01& $-$00&  +01& +03  \\
63 &  Eu~{\sc ii} & 1356 &  0.51 &  2 &  +1.28& (  1)& +0.40& (  2)& +1.07& (  1)&    +15& +10&  $-$04& $-$01&  +02& +04  \\
\hline
\end{tabular}
\end{center}
\scriptsize
(1) Atomic number. (2) Element species. (3) 50\% condensation temperature (in K) taken from Table~8 of 
Lodders (2003). (4) Reference solar abundances (in the usual normalization of H = 12.00) taken 
from Anders \& Grevesse's (1989) compilation, except that 7.50 is adopted for Fe  and the solar 
system abundance of 3.28 (Lodders 2003) is used for Li corresponding to the young Sun. 
(5) Number of lines adopted for this species.
(6) Mean of [X/H]$_{\rm A}$ (relative abundance for A in comparison with the Sun) 
averaged over lines. (7) Actual number of lines used for deriving $\langle$[X/H]$_{\rm A}\rangle$.
(8) Mean of [X/H]$_{\rm B}$. (9) Actual number of lines employed for $\langle$[X/H]$_{\rm B}\rangle$.
(10) Mean of $\Delta A^{\rm X}_{{\rm A}-{\rm B}}$ (line-by-line differential abundance between
A and B) averaged over lines. (11) Actual number of lines used for calculating
$\langle \Delta A^{\rm X}_{\rm A-B}\rangle$. 
(12) Abundance change for A in response to $\Delta T_{\rm eff} = +300$~K.  
(13) Abundance change for B in response to $\Delta T_{\rm eff} = +300$~K.   
(14) Abundance change for A in response to $\Delta v_{\rm t} = +0.3$~km~s$^{-1}$.  
(15) Abundance change for B in response to $\Delta v_{\rm t} = +0.3$~km~s$^{-1}$.
(16) Abundance change for A in response to $\Delta \log g = +0.1$~dex.  
(17) Abundance change for B in response to $\Delta \log g = +0.1$~dex. 
All abundance-related data ($A_{\odot}$, $\langle$[X/H]$\rangle$, $\Delta$) are 
in unit of dex.  Note that only the 1st and 2nd decimals are shown 
in the data of (12)--(17) (i.e., they should be divided by 100). 
\end{minipage}
\end{table*}

\section{Discussion}

\subsection{Impact of new $T_{\rm eff}$ and $v_{\rm t}$}

We first discuss the results of $T_{\rm eff}$ and $v_{\rm t}$, 
which were spectroscopically established based on Fe lines (Sect.~3), 
in comparison with those employed in the past studies (cf. Table~1). 

Regarding the effective temperature, since previous papers reported values 
in the ranges of $T_{\rm eff,A} \sim$~7600--8200~K and 
$T_{\rm eff,B} \sim$~6900--7600~K, our results of 7990~K (A) and 7570~K (B) 
rather belong to the high-scale group among the literature values.

As to the microturbulence, our $v_{\rm t,A}$ of 3.5~km~s$^{-1}$ corresponds to 
the low-scale group among the published values ($\sim$~3--7~km~s$^{-1}$),
while $v_{\rm t,B}$ of 2.6~km~s$^{-1}$ is almost the mean of previous 
determinations (2.1--3.2~km~s$^{-1}$)

In order to demonstrate that abundance determinations may be 
significantly affected by $T_{\rm eff}$ and $v_{\rm t}$, we compare 
the adopted parameters as well as the resulting metallicity in Paper~I
($T_{\rm eff.A}$/$T_{\rm eff.B}$, $v_{\rm t,A}$/$v_{\rm t,B}$ and 
[Fe/H]$_{\rm A}$/[Fe/H]$_{\rm B}$) = (7570/6980, 3.8/3.1, $+0.11/-0.41$)
with those derived in this study (7990/7570, 3.5/2.6, $+0.37/-0.01$).
The reason why the new [Fe/H] values in this investigation have appreciably 
increased by $\sim$~0.3--0.4~dex for both A and B in comparison with Paper~I 
is due to the increase in $T_{\rm eff}$ by $\sim$~400--600~K and to
the decrease in $v_{\rm t}$ by $\sim$~0.3--0.5~km~s$^{-1}$, because
both act in the direction of increasing $A$(Fe) as can be seen in columns 12--15
of Table~5).\footnote{It was mentioned in Sect.~4.3 that the effect of 
changing $v_{\rm t}$ on the abundances is not so significant (in contrast 
to the case of changing $T_{\rm eff}$). It should be noted, however, that this is due to 
the specific situation in this study: (i) a large number of weak Fe lines (for which 
the resulting abundances hardly depend upon $v_{\rm t}$) could be employed and (ii) strong 
lines satisfying $10^{6}W_{\lambda}/\lambda > 30$ were excluded in deriving the mean abundance.
Generally, changing $v_{\rm t}$ (even by only a few tens km~s$^{-1}$) may bring about
appreciable variations in the resulting abundances depending on the lines used
in the analysis.}  
 
\subsection{Trends of chemical abundances}

Based on the results in Table~5, 
$\langle$[X/H]$_{\rm A}\rangle$, $\langle$[X/H]$_{\rm B}\rangle$, and 
$\langle \Delta A^{\rm X}_{\rm A-B}\rangle$ are plotted against $Z$ in 
Fig.~8a, 8b, and 8c, respectively. Considering the impact of uncertainties 
in atmospheric parameters (cf. Table~3 and Table~5) and the size of standard 
deviations in the averages (cf. ``abundances.txt''), typical statistical errors 
involved with the data symbols in Fig.~8 may be estimated as $\pm \lesssim$~0.1--0.2~dex. 
While some additional systematic errors can not be ruled out in [X/H] values 
related to other factors (e.g., reference solar abundances, $gf$ values, 
non-LTE effect for $Z\ge 21$ elements, etc.), $\Delta A^{\rm X}_{\rm A-B}$ values 
(line-by-line differential abundances) should be almost irrelevant to such 
systematic error sources because of being almost canceled out.

Several notable features are observed by inspecting Fig.~8 as summarized below 
(the symbols ``$\langle$'' and  ``$\rangle$'' to indicate average values are 
omitted for simplicity):
\begin{itemize}
\item
Roughly speaking, [X/H] (departure from the solar composition) 
tends to increase with $Z$ in the global sense for both A and B (Fig.~8a and 8b); 
i.e., [X/H]$_{\rm A}$~$<0$ at $Z \lesssim 10$, [X/H]$_{\rm A}$~$\sim 0$ at $10 \lesssim Z \lesssim 20$, 
and [X/H]$_{\rm A}$~$>0$ at $Z \gtrsim 20$ for RR~Lyn~A; while [X/H]$_{\rm B}$~$<0$ at $Z \lesssim 10$, 
[X/H]$_{\rm B}$~$\sim 0$ at $10 \lesssim Z \lesssim 30$, and [X/H]$_{\rm B}$~$>0$ at $Z \gtrsim 30$
for RR~Lyn B. Since the gradient of this trend is steeper for A than for B, 
$\Delta A^{\rm X}_{\rm A-B}$ also shows a similar $Z$-dependent tendency (Fig.~8c)
\item
We have confirmed that A is an Am star, which shows typical characteristics of 
Am phenomenon (e.g., underabundances of CNO, moderate enrichment of Fe-group elements, 
and large overabundances of s-process elements and rare earths). However, the drastic 
depletion of Sc ([Sc/H]$_{\rm A} \sim -1.6$) is worth noting, which is rather puzzling
because the deficiency of Ca (considered to conform with Sc in Am stars) is only mild
([Ca/H]$_{\rm A} \sim -0.3$). This problem is separately discussed in Sect.~5.4.
\item
In contrast, chemical aberrations of B are only marginal: rather tight and gradual 
$Z$-dependence of [X/H]$_{\rm B}$, while [X/H]~$\sim 0$ for elements of intermediate 
$Z$ (such as Fe group). Therefore, B may be classified as a weak Am star of almost 
solar metallicity. 
\end{itemize} 

\subsection{RR~Lyn~B is not metal-deficient}

While the conclusion of metal-rich nature for RR~Lyn~A is more or less consistent 
with most of the past publications, our finding that  ``RR~Lyn B is a solar-metallicity
star ([Fe/H]$_{\rm B} \sim 0$) with a gradually increasing trend of [X/H]$_{\rm B}$ 
with $Z$'' is significant, because it apparently conflicts with the results of previous 
investigations, most of which concluded B to be of subsolar metallicity (cf. Table~1). 
\begin{itemize}
\item
Kondo (1976) mentioned that Fe is likely to be deficient in B.
\item
Lyubimkov \& Rachkovskaya's (1995b) separate analysis for A and B (based on 
CCD spectra) yielded [Fe/H]$_{\rm B}$ = $-0.33$, from which they concluded 
that B may be a $\lambda$~Boo star of metal-deficient peculiarity.  
\item
Burkhart \& Coupry (1991) reported [Fe/H]$_{\rm B}$ = $-0.3$, though
details of their analysis (e.g., $v_{\rm t}$, $\log g$) are not described. 
\item
Hui-Bon-Hoa (2000) derived [Fe/H]$_{\rm B}$ = $-0.19$, though with a remark 
of ``uncertain value''. 
\item
Khaliullin et al. (2001) obtained [Fe/H]$_{\rm B}$ = $-0.24 (\pm 0.06)$,
though this is a photometric metallicity derived from the length of the 
blanketing vector in two-color diagram.
\item
Jeong et al. (2017) concluded rather confidently based on their abundance 
determinations that B ([Fe/H]$_{\rm B}$ = $-0.33$) is a $\lambda$~Boo star, 
because they observed characteristic abundance differences between 
volatile (low condensation temperature $T_{\rm c}$) and refractory 
(high $T_{\rm c}$) elements; i.e., they found that the former is almost 
solar but the latter is appreciably deficient (cf. their Fig.~4B).
\end{itemize}

Since the discrepancy of [Fe/H]$_{\rm B}$ between Paper~I ($-0.41$) and
this study ($-0.01$) can be reasonably explained by the difference in
$T_{\rm eff}$ and $v_{\rm t}$ (cf. Sect.~5.1), the atmospheric parameters 
adopted by these authors were examined. It was then found that the values 
of $T_{\rm eff,B}$ adopted in most of these papers are appreciably lower 
than ours (7570~K) by $\sim$~400--700~K: 6900~K (Kondo 1976), 
7150~K (Lyubimkov \& Rachkovskaya 1995b), 6980~K (Khaliullin et al. 2001), 
7210~K (Jeong et al. 2017); though Burkhart \& Coupry (1991) and 
Hui-Bon-Hoa (2000) employed $\sim$~7500--7600~K like this study. 
Therefore, in view of the significant role of $T_{\rm eff}$  
in abundance derivation (cf. Sect.~5.1), we suspect that the 
apparently subsolar metallicity of RR~Lyn~B concluded by these 
authors may be mainly due to their use of inadequately low $T_{\rm eff}$.

The systematic trend of [X/H]$_{\rm B}$ in terms of $T_{\rm c}$,
which was reported by Jeong et al. (2017) as an evidence for a 
$\lambda$~Boo star, is not observed in our analysis.
This can be recognized in Fig.~9, where our abundance 
results in Table~5 are plotted against $T_{\rm c}$.
It is apparent that such a $T_{\rm c}$-dependent trend in [X/H]$_{\rm B}$
with a negative slope as found by Jeong et al. (2017; cf. their Fig.~4B) 
is absent in our [X/H]$_{\rm B}$ results (cf. Fig.~9b; the trend is even 
the contrary to their claim, because the slope is rather positive).
Although several factors may be involved with this disagreement
(e.g., difference in atmospheric parameters, their neglect of non-LTE effect
for volatile elements, etc.), much can not be said unfortunately, because 
sufficient details of the spectral lines they used are not presented.

\begin{figure}[H]
\begin{minipage}{70mm}
\begin{center}
  \includegraphics[width=7.0cm]{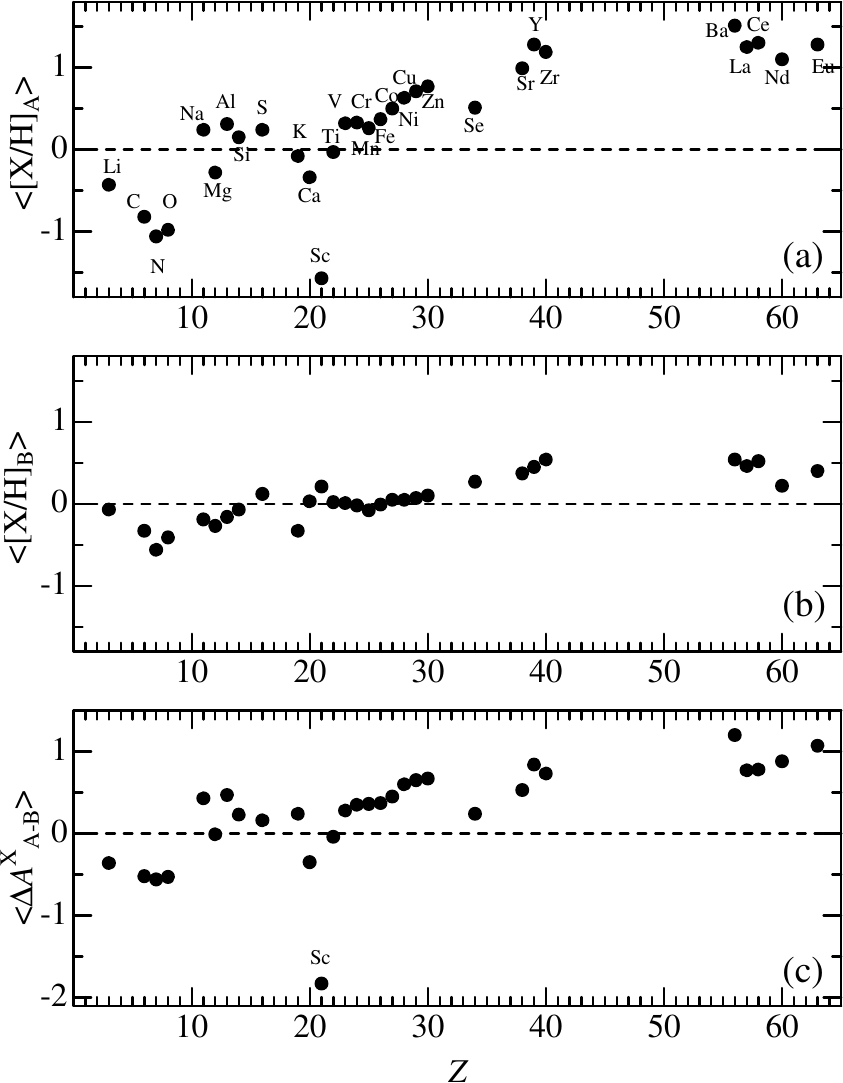}
\end{center}
   \caption{
Plotted against $Z$ (atomic number) are  
(a) $\langle$[X/H]$_{\rm A}\rangle$ (averaged 
differential abundance of element X relative to the Sun for RR~Lyn~A), 
(b) $\langle$[X/H]$_{\rm B}\rangle$  (ditto for RR~Lyn~B), and 
(c) $\langle \Delta A^{\rm X}_{{\rm A}-{\rm B}}\rangle$ (averaged 
line-by-line differential abundance of element X between A and B), 
for 30 elements based on the data in Table~5. Regarding the elements 
where two results of different ionization stages are available 
(Mg, Al, Si, Ca, Ti, Cr, Mn, Fe), 
their mean values are adopted here (exceptionally, the result for 
V~{\sc ii} is used for V because that for V~{\sc i} based on only 
one line is appreciably discrepant).
} 
   \label{Fig8}
\end{minipage}
\end{figure}

\begin{figure}[H]
\begin{minipage}{70mm}
\begin{center}
  \includegraphics[width=7.0cm]{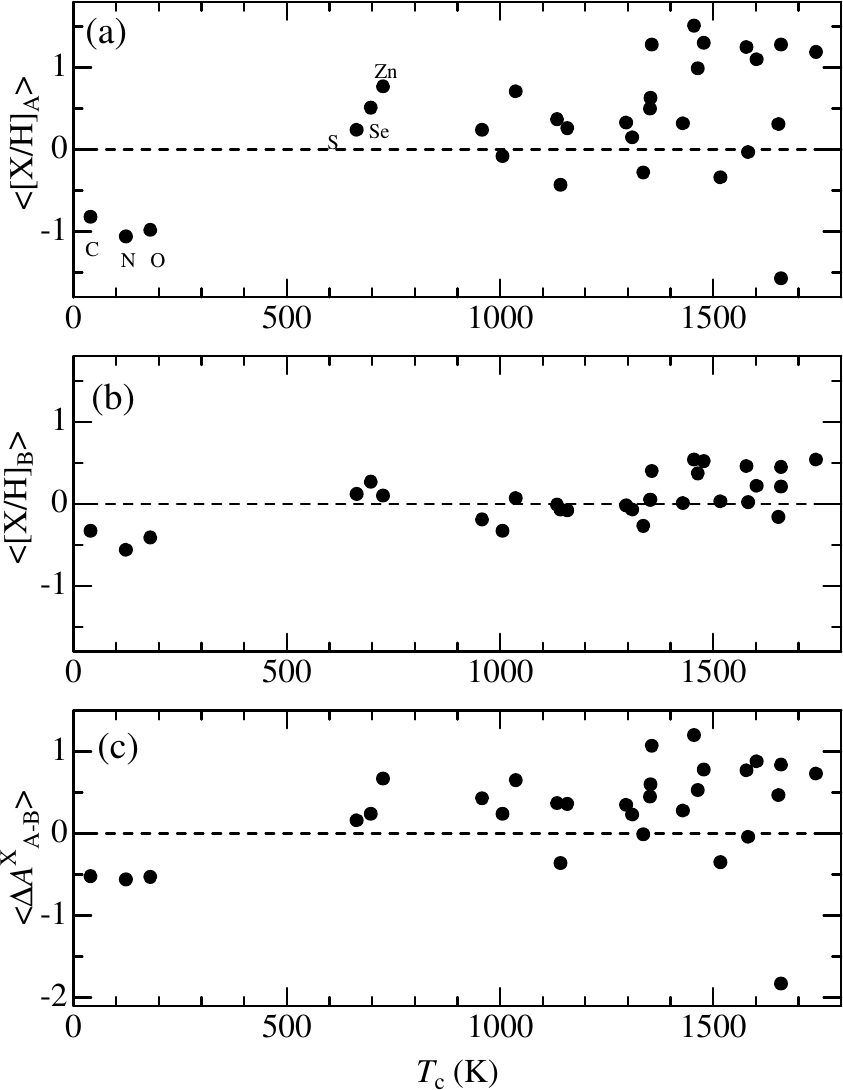}
\end{center}
   \caption{
The results of   
(a) $\langle$[X/H]$_{\rm A}\rangle$, (b) $\langle$[X/H]$_{\rm B}\rangle$, and 
(c) $\langle \Delta A^{\rm X}_{{\rm A}-{\rm B}}\rangle$ derived for 30 elements 
are plotted against $T_{\rm c}$ (50\% condensation temperature). 
Otherwise, the same as in Fig.~8.   
} 
   \label{Fig9}
\end{minipage}
\end{figure}

\subsection{Scandium problem in RR~Lyn~A}

It is remarkable that Sc ($Z = 21$) shows a conspicuously large underabundance 
([Sc/H] = $-1.6$) in RR~Lyn~A, which markedly deviates from the global trend 
of [X/H] distributing around $\sim 0$ (within $\pm$ several tenths dex) for 
$10 \lesssim Z \lesssim 25$ elements (Fig.~8a).
It is worth noting that Ca ($Z = 20$) is only mildly subsolar (by $\sim -0.3$~dex) 
and does not conform to the substantial deficiency of Sc. 
This is somewhat strange, because Ca and Sc are considered to behave similarly 
in Am stars, where weak Ca~{\sc ii} and/or Sc~{\sc ii} lines (along with 
enhanced strengths of heavy metal lines) are their discovery criteria in 
spectral classifications (cf. Table~1 in Preston 1974). 
Actually, according to Fig.~5 of Smith's (1996) review paper, published results 
of [Ca/H] and [Sc/H] appear to be rather similar in Am--Fm stars. 

In order to examine whether such a striking abundance difference between Sc 
and Ca could be possible in Am stars, Sc abundances of 101 (mostly 
A-type) stars studied by Takeda et al. (2018) were determined from the 
Sc~{\sc ii}~5526 line, and compared them with the Ca abundances derived from 
the Ca~{\sc i} 6162 line. This supplementary analysis is separately described 
in Appendix~A. 

It was then found as the characteristic trend of Am stars that 
(i) [Sc/H] and [Ca/H] positively correlate with each other and 
(ii) the gradient d[Sc/H]/d[Ca/H] is greater than 1 (Fig.~10c),
which means that the extent of deficiency is larger for Sc
than for Ca ($|[{\rm Sc/H}]| \gtrsim |[{\rm Ca/H}]|$).
Although the ratio of [Sc/H]/[Ca/H]~$\simeq$~4--5 found for RR~Lyn~A appears 
somewhat too large compared to other Am stars (typically around $\sim 2$),
it might be regarded as a result of random scatter because the correlation 
becomes progressively diversified as the deficiency is enhanced.
Accordingly, we may state that the markedly large discordance between 
[Sc/H] and [Ca/H] observed in RR~Lyn~A is qualitatively understandable 
within the framework of the general characteristics of Am stars, though 
it is a rather rare phenomenon in the quantitative sense.
 
From the viewpoint of the theoretical side, a number of researchers have
devoted efforts to elucidating the cause of Am phenomenon over the past 
half century. Above all, chemical segregation (element diffusion) coupled with 
mass loss, which may take place in the stable atmosphere/envelope has been 
regarded as a promising mechanism, though the abundance behaviors Ca and Sc 
in Am stars are not yet successfully explained by the diffusion theory
(see Michaud et al. 2015; and the references therein). 
Alternatively, some authors considered another possibility for the origin 
of Am anomalies, such as due to accretion of interstellar matter coupled 
with charge-exchange reactions (B\"{o}hm-Vitense 2006; Jeong et al. 2017).

In any event, these remarkable features revealed from this study 
(i.e., very discordant abundance deficiency between Sc and Ca in RR~Lyn A, 
while almost normal Ca and even somewhat supersolar Sc in RR~Lyn~B 
of lower $T_{\rm eff}$ by only $\sim$~400~K) would serve as a valuable 
observational constraint and an ideal testing bench for any theory challenging 
to explain the chemical peculiarities in Am stars, 
because stellar parameters (including the age of $\sim 1\times 10^{9}$~yr; 
Khaliullin et al. 2001) are well established for both components.  

\section{Summary and conclusion}

RR~Lyn A+B is an eclipsing binary system (with orbital period of $\sim 10$~d) 
consisting of two main-sequence stars, in which the hotter A is classified 
as an metallic-line A-type star. Several previous investigations reported 
that surface abundances of these two components are considerably different 
from each other (i.e., A is metal-rich but B is even metal-poor) despite that 
both should have born with the same chemical composition. 

Therefore, this system may be regarded as an important key to clarify the 
physical mechanism triggering chemical peculiarities (Am phenomenon) in the stellar 
surface by carefully examining the detailed abundances of various elements 
for both components.

However, extensive elemental abundance study for both A and B based on 
high-quality data are still insufficient, reflecting the difficulty of 
analyzing complex double-lined spectra of RR~Lyn.  
Besides, atmospheric parameters (especially $T_{\rm eff}$ and $v_{\rm t}$), 
which may significantly affect abundance determinations, are not yet  
reliably established, because appreciably diversified values have been 
reported in the past studies.

Motivated by this situation, the purpose of this study was to carry out 
a detailed spectroscopic study of RR~Lyn A+B in order to determine 
the key atmospheric parameters and elemental abundances of A and B 
as precisely as possible and to examine how they compare with each other.

Regarding the basic observational data, the spectrum-disentangle technique
was applied to a set of double-line spectra taken at different orbital 
phases to obtain the decomposed spectra of A and B. Based on these disentangled 
spectra (covering 3900--9200~\AA), many lines judged to be usable were identified 
and their $W_{\lambda}$ were measured by the direct Gaussian-fitting. 
In addition, $W_{\lambda}$ values of important line features (even if they are 
blended with other lines or consist of complex multi-components) were evaluated 
by applying the spectrum-synthesis technique.
 
The values of ($T_{\rm eff}$, $v_{\rm t}$) were determined from many Fe~{\sc i} 
lines by requiring that abundances do not show any systematic dependence upon
$W_{\lambda}$ and $\chi_{\rm low}$, where stronger lines  
($10^{6} W_{\lambda}/\lambda > 30$) had to be discarded because they turned out 
to behave anomalously. This analysis yielded (7990~K, 3.5~km~s$^{-1}$) 
and (7570~K, 2.6~km~s$^{-1}$) for A and B, respectively. 

The chemical abundances of 30 elements (39 species) were derived from 
the $W_{\lambda}$ values of many lines, where the non-LTE effect was taken 
into account for comparatively lighter elements ($Z \le 20$ and $Z = 30$).
The following characteristics were found in the resulting [X/H]$_{\rm A}$ 
and [X/H]$_{\rm B}$. 
\begin{itemize}
\item[--]
Roughly speaking, [X/H] tends to increase with $Z$ for both A and B.
That is, light elements ($Z < 10$) are subsolar, intermediate elements 
($10 \lesssim Z \lesssim$~xx) are nearly solar; and heavy elements 
(xx~$\lesssim Z$) are supersolar, where xx~$\sim 20$ for A and xx~$\sim 30$ for B.
These are the characteristic trends observed in Am stars.
\item[--]
Abundance peculiarities are quantitatively more conspicuous in A than in B,   
resulting in a similar increasing trend with $Z$ also for the differential 
abundances between A and B ($\Delta A^{\rm X}_{{\rm A}-{\rm B}}$).
\item[--]
Yet, regarding the hotter A, Sc shows a strikingly large deficiency 
([Sc/H]$_{\rm A} \simeq -1.6$) deviating the global trend.
Although it is somewhat strange that Ca is only mildly underabundant 
([Ca/H]$_{\rm A} \sim -0.3$) and does not conform to Sc (since both are 
considered to behave similarly), such a large discordance may not necessarily 
be improbable, since Sc tends to be more deficient than Ca in Am stars. 
\item[--]
It is significant that the abundances of iron group elements 
($20 \lesssim Z \lesssim 30$) in RR~Lyn B are almost solar ([Fe/H]$_{\rm B} \simeq 0.0$).
This conclusion is in contrast with the results of previous studies, 
most of which reported that B is metal-deficient (such as like $\lambda$~Boo stars) 
by several tenths dex. This discrepancy may be attributed to the difference 
in the adopted atmospheric parameters (e.g., $T_{\rm eff}$). 
\end{itemize} 

In short, RR~Lyn~A is a typical (classical) Am star of supersolar metallicity
([Fe/H]$_{\rm A} \simeq +0.4$) with a conspicuously large Sc deficiency. 
Meanwhile, RR~Lyn~B is a solar-metallicity star ([Fe/H]$_{\rm B} \simeq 0.0$) 
with a trend of [X/H]$_{\rm B}$ gradually increasing with $Z$ (i.e., a 
near-normal star accompanied by weak Am characteristics).
It is thus important to elucidate the reason why such dissimilar types of 
chemical anomalies are observed in A and B (with a small $T_{\rm eff}$ 
difference of $\sim$~400~K), for which challenges of theoreticians are awaited.  

\appendix
\section{Sc and Ca abundances of A-type stars}

In view of the remarkably large underabundance of Sc found in RR~Lyn~A 
($\sim -1.6$~dex) despite of only a mild deficiency ($\sim -0.3$~dex) 
of Ca (which is considered to behave along with Sc in Am stars), it is worth
examining the trends of Sc and Ca abundances and their mutual correlations
in a large sample of A-type stars, in order to see whether  
such a substantial difference between Sc and Ca is possible in Am stars. 

Conveniently, Takeda et al. (2018) once carried out CNO abundance 
determinations for 101 main-sequence stars (in the parameter ranges 
of $0 \lesssim v_{\rm e}\sin i \lesssim 100$~km~s$^{-1}$ and  
$7000 \lesssim T_{\rm eff} \lesssim 10000$~K), most of which are 
A-type stars including 25 Am stars (i.e., classified as Am in Hipparcos 
catalogue; cf. Sect.~2 in Takeda et al. 2018). 
Therefore, Sc abundances of these 101 stars were determined (in LTE)
by applying the spectrum-fitting technique to the wavelength region
comprising the Sc~{\sc ii} 5526.79 line ($\chi_{\rm low} = 1.768$~eV,
$\log gf = +0.024$) based on the same observational data and 
the same atmospheric parameters as adopted therein.

Although abundance solutions were successfully converged for most cases 
(excepting 5 stars), a possibility of converging at a spurious solution
still remains, especially for broad-line cases of larger $v_{\rm e}\sin i$.
Therefore, the credibility of the resulting abundance was assessed as follows.   
The equivalent width of the Sc~{\sc ii} 5526 line ($W_{5526}$) was 
calculated from the fitting-based abundance as done in Sect.~4.1.
Meanwhile, the critical line strength at the detectability limit may 
be roughly estimated as $W_{*} \simeq h \times \epsilon$, where $h$ is 
the full-width at the half maximum of the line (evaluated from 
$v_{\rm e}\sin i$) and $\epsilon$ is the minimum-detectable line-depth 
depending upon SNR (signal-to-noise ratio) as $\epsilon \simeq 1/{\rm SNR}$.
If $W_{5526}$ was smaller than $W_{*}$, then the abundance was judged to 
be unreliable. Besides, for the unsuccessful cases where fitting solutions 
failed to converge, the upper-limit abundances ware derived from $W_{*}$.

The Ca abundances (in LTE) of 101 stars were actually derived (though the 
results were not published) in the study of Takeda et al. (2018) by the 
fitting analysis of the 6146--6163~\AA\ region (see Fig.~4 therein) including 
the Ca~{\sc i} 6162.16 line ($\chi_{\rm low} = 1.899$~eV, $\log gf = +0.100$). 
The reliabilities of these Ca abundances were assessed in the same manner 
as done for Sc.

Thus obtained Sc and Ca abundances of 101 stars are summarized in ``scabund101.txt'' 
of the online material. Their dependence upon $T_{\rm eff}$ or 
$v_{\rm e}\sin i$ as well as mutual correlations are illustrated in Fig.~10,
where the spectrum fitting in the Sc~{\sc ii} 5526 line region for two 
representative stars (Am star HD~072037 and non-Am star HD~189849) is also 
demonstrated (Fig.~10f). 
In addition, since Mashonkina \& Fadeyev (2024) recently determined Ca and 
Sc abundances for a number of  A-type stars by taking into account the non-LTE effect.
their non-LTE $A$(Ca) and $A$(Sc) values (taken from their Table~1 and Table~2) 
are also overplotted in this figure for comparison. 
The following characteristics are observed from this figure regarding the Sc 
and Ca abundances of Am stars.
\begin{itemize}
\item
The abundances of both elements are generally subsolar in Am stars 
by several tenths to $\sim 1$~dex, though the dispersion for Sc  
($-1.5 \lesssim {\rm [Sc/H]} \lesssim 0$) is somewhat larger than that 
for Ca ($-1 \lesssim {\rm [Ca/H]} \lesssim 0$) (Fig.~10a and 10d). 
\item
A weak $v_{\rm e}\sin i$-dependent trend is seen for both Sc and Ca 
in Am stars, in the sense that slower rotators tend to show larger 
deficiencies (Fig.~10b and 10e). 
\item
$A$(Sc) and $A$(Ca) are positively correlated with each other, though 
the dispersion tends to be larger with the d$A$(Sc)/d$A$(Ca) gradient  
being steeper ($> 1$) towards increasing the deficiency in Am stars 
(Fig.~10c).
\item
The non-LTE results of $A$(Ca) and $A$(Sc) derived by Mashonkina \& Fadeyev 
(2024) are almost in accord with the present LTE results as seen from 
Fig.~10a--10e. Though their $A$(Ca) values tend to be somewhat higher 
in the strict sense, the same overall trends are observed.  
\end{itemize}

\begin{figure}[H]
\begin{minipage}{75mm}
\begin{center}
  \includegraphics[width=7.5cm]{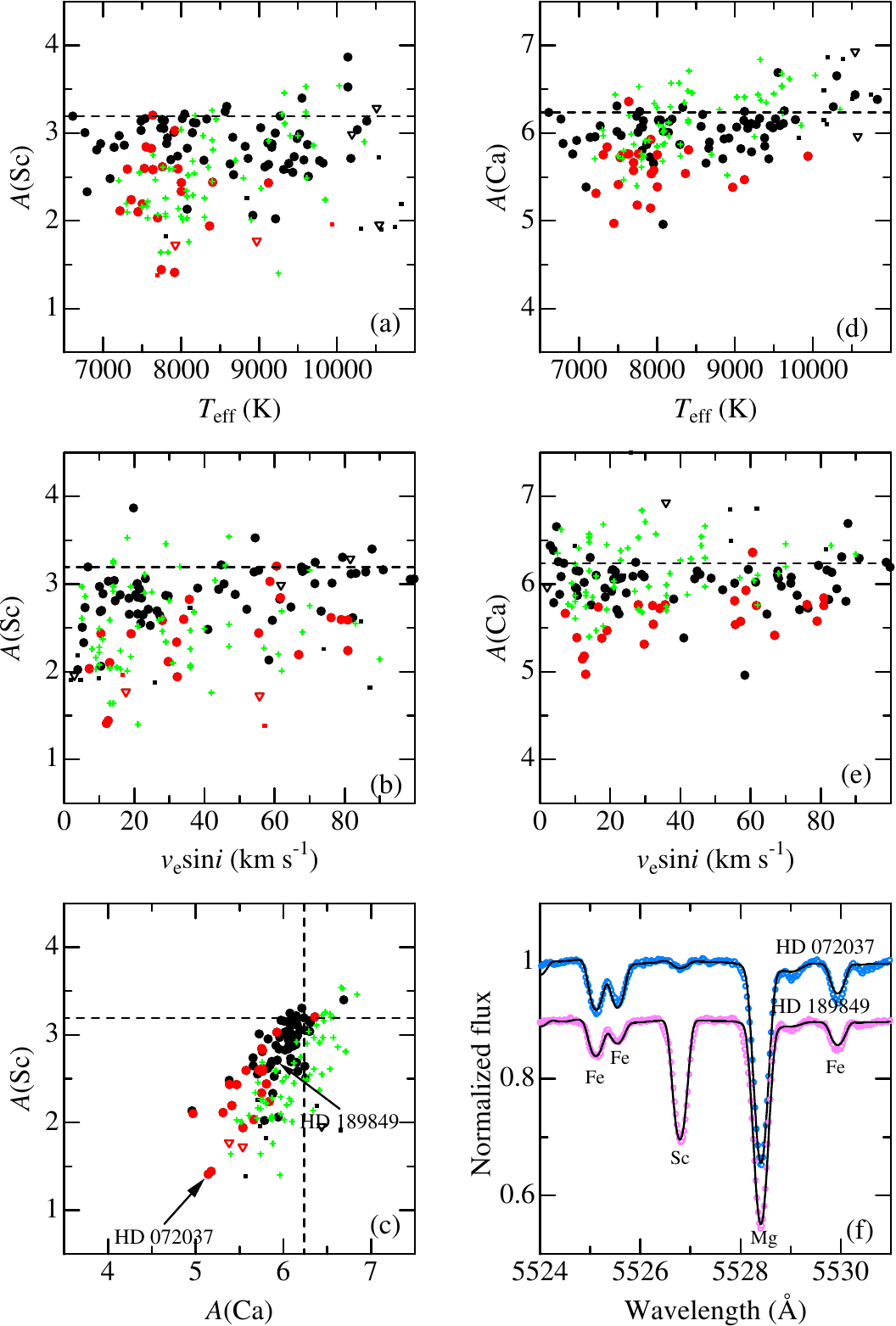}
\end{center}
   \caption{Trends and correlations of Sc and Ca abundances determined 
for 101 sharp-lined (mostly) A-type stars. In panels (a), (b), (d), and (e) 
are plotted the abundances of $A$(Sc) (left) and $A$(Ca) (right) 
against $T_{\rm eff}$ (top) and $v_{\rm e}\sin i$ (middle).
The correlation between $A$(Sc) and $A$(Ca) is illustrated in panel (c).
Panel (f) demonstrates the spectrum fitting in the region
comprising the Sc~{\sc ii} 5526 line for two representative cases 
of HD~072037 and HD~189849 (depicted in the same manner as in Fig.~7).
In each panel showing $A$(Sc) or $A$(Ca), ordinary data, unreliable data,
and upper limits (indeterminable cases) are 
denoted by filled circles, small dots, and open downward triangles,
respectively. The data for Am stars (25 out of 101 stars) are highlighted 
in red. The abundances of the reference star Procyon (HD~061421)
(3.19 for Sc and 6.24 for Ca; which may be regarded as good proxies for 
the solar abundances) are indicated by dashed lines.  
For comparison, the non-LTE Ca and Sc abundances determined by 
Mashonkina \& Fadeyev (2024), which were taken from their Table~1 and Table~2,
are also overplotted by light-green crosses in panels (a)--(e).
} 
   \label{Fig10}
\end{minipage}
\end{figure}

\section*{Acknowledgements}
\vspace{-1em}
Data analysis (numerical calculation of spectrum disentangling) was carried 
out on the Multi-wavelength Data Analysis System operated by the Astronomy 
Data Center (ADC), National Astronomical Observatory of Japan.
This investigation has made use the VALD database operated at Uppsala University,
the Institute of Astronomy RAS in Moskow, and the University of Vienna.

\section*{Online materials}

This article accompanies the following online materials 
(electronic data files).
See ``ReadMe.txt'' for the details about their contents.
\begin{itemize}
\item
ReadMe.txt 
\item
specdata\_A.txt 
\item
specdata\_B.txt
\item
ewlines\_A.txt 
\item
ewlines\_B.txt
\item
abundances.txt 
\item
scabund101.txt
\end{itemize}



\end{document}